\documentclass{aa}
\usepackage[varg]{txfonts}
\usepackage[]{natbib}
\bibpunct{(}{)}{;}{a}{}{,} 

\newcommand{\msun}{{M$_\odot$}}
\newcommand{\Al}{$^{26}$Al}

\newcommand{\Ne}{$^{22}$Ne}
\newcommand{\F}{$^{19}$F}
\newcommand{\Cl}{$^{36}$Cl}
\newcommand{\Ca}{$^{41}$Ca}

\newcommand{\Aov}{$\alpha_{\rm ov}$}
\newcommand{\Denv}{log(D$_{\rm env}$/cm$^{2}$s$^{-1}$)}

\begin{document}

\title{The impact of asteroseismically calibrated internal mixing on nucleosynthetic wind yields of massive stars}
\titlerunning{Nucleosynthesis and Asteroseismology}
\author{Hannah E. Brinkman \inst{1}
\and Lorenzo Roberti \inst{2,3,4}
\and Alex Kemp \inst{1}
\and Mathias Michielsen \inst{1}
  \and Andrew Tkachenko\inst{1}
     \and Conny Aerts \inst{1,5,6}}

\offprints{hannah.brinkman@kuleuven.be}

\institute{Institute of Astronomy, KU Leuven, Celestijnenlaan 200D, 3001, Leuven, Belgium \and Konkoly Observatory, Research Centre for Astronomy and Earth Sciences, HUN-REN, Konkoly Thege Mikl\'{o}s \'{u}t 15-17, H-1121 Budapest, Hungary \and CSFK, MTA Centre of Excellence, Budapest, Konkoly Thege Miklós út 15-17, H-1121, Hungary \and INAF -- Osservatorio Astronomico di Roma Via Frascati 33, I-00040, Monteporzio Catone, Italy \and Department of Astrophysics, IMAPP, Radboud University Nijmegen, PO Box 9010, 6500 GL Nijmegen, The Netherlands \and Max Planck Institute for Astronomy, Königstuhl 17, 69117 Heidelberg, Germany
}

\date{Received February 16, 2024; Revised April 26, 2024; Accepted May 10, 2024}

\abstract{Asteroseismology gives us the opportunity to look inside stars and determine their internal properties, such as the radius and mass of the convective core. Based on these observations, estimations can be made for the amount of the convective boundary mixing and envelope mixing of such stars, and the shape of the mixing profile in the envelope. However, these results are not typically included in stellar evolution models.}
{We aim to investigate the impact of varying convective boundary mixing and envelope mixing in a range based on asteroseismic modelling in stellar models up to the core-collapse, both for the stellar structure and for the nucleosynthetic yields. In this first study, we focus on the pre-explosive evolution and evolve the models to the final phases of carbon burning. This set of models is the first to implement envelope mixing based on internal gravity waves for the entire evolution of the star.}
{We use the MESA stellar evolution code to simulate stellar models with an initial mass of 20 \msun{} from the ZAMS up to a central core temperature of 10$^{9}$ K, which corresponds to the final phases of carbon burning. We vary the convective boundary mixing, implemented as "step-overshoot", with the overshoot parameter (\Aov{}) in the range 0.05 - 0.4. We vary the amount of envelope mixing (\Denv{}) in the range 0-6 with a mixing profile based on internal gravity waves. To study the nucleosynthesis taking place in these stars in great detail, we use a large nuclear network of 212 isotopes from $^{1}$H to $^{66}$Zn.}
{Enhanced mixing according to asteroseismology of main-sequence stars, both at the convective core boundary and in the envelope, has significant effects on the nucleosynthetic wind yields. This is especially the case for \Cl{} and \Ca{}, whose wind yields increase by ten orders of magnitude compared to those of the models without enhance envelope mixing. Our evolutionary models beyond the main sequence diverge in yields from models based on rotational mixing, having longer helium burning lifetimes and lighter helium-depleted cores.}
{We find that the asteroseismic ranges of internal mixing calibrated from core hydrogen burning stars lead to similar wind yields as those resulting from the theory of rotational mixing. Adopting the seismic mixing levels beyond the main sequence, we find earlier transitions to radiative carbon burning compared to models based on rotational mixing because they have lower envelope mixing in that phase. This influences the compactness and the occurrence of shell-mergers, which may affect the supernova properties and explosive nucleosynthesis.}

\maketitle
\section{Introduction} \label{intro}
Massive stars (M$_{*}$ $\geq$ 10 \msun{}, in this context) are one of the main sources of nucleosynthesis in the Universe \citep[see, e.g.,][and many others]{Hirschi2005, Ekstrom2012, Pignatari2016, Ritter2018, LandC2018}. During their lives and deaths they turn the lighter elements into heavier ones. The nucleosynthetic yields of these stars, aside from the initial mass and initial metallicity, depend on three main ingredients; (i) the nuclear reaction rates, which govern the production/destruction of isotopes inside the star \citep[see, e.g.,][]{Kobayashi2020}, (ii) the internal mixing, which is responsible for moving the isotopes from where they are produced to other layers of the star \citep{Pedersen2021}, and (iii) mass loss, through stellar winds \citep{KudritzkiPuls2000, LangerReview, Smith2014Review}, supernova explosions (including the explosion mechanism and the isotopes produced during the explosive nucleosynthesis) \citep{WoosleyHeger2002, Nomoto2013}, and binary interactions \citep{Sana2012}, which expel the isotopes into the interstellar medium. In this work, we focus on the effects of internal mixing on the nucleosynthetic wind yields, while we keep the nuclear reaction rates and the mass loss rates to standard values, and we do not consider the supernova explosion. We do this because even though the internal mixing is very important for the wind yields, it is currently not well constrained. The internal mixing near the convective core during the main sequence is important because it impacts the mass of the hydrogen-depleted core at the end of this stage, setting the stage for the nuclear burning in the more evolved stages \citep{Johnston2021,Pedersen2022}, and how much of the burning products reach the surface of the star from where they are expelled.\\
\indent The main mixing mechanism for massive stars is convective mixing. It transports both the nuclear energy produced in the burning zones (either the stellar core or burning shells), as well as the isotopes produced in these areas. However, since convection is a three-dimensional process (while stellar evolution on nuclear timescales can only be handled in 1D), it can only be modelled by a parametric approach. Aside from the choice between the Ledoux or Schwarzschild criterion for convection, it needs to be taken into account that the convective motions do not stop at the boundaries of such zones. Convective boundary mixing (CBM hereafter) is needed to correct for this, and  with the introduction of a treatment of CBM, free parameters are also introduced. Commonly these parameters are either f$_{ov}$ or $\alpha_{ov}$ \citep[with a mapping factor of \Aov{}/f$_{ov}$ $\sim$ 10, see, e.g.,][Sec. 3.1]{Kaiser2020}{}{}. f$_{ov}$ is linked to ``exponential overshoot'', where the CBM is treated via an exponential decay of the convective diffusion coefficient beyond the convective boundary as given in \citet{Herwig1997};
\begin{equation}{
    D_{ov} = D_{0}\cdot exp(-2z/f_{ov} \cdot H_{p})
    }
\end{equation}
Here, the diffusion coefficient, D$_{ov}$, is a function of the distance from the convective boundary, the free parameter f$_{ov}$, and the pressure scale height, H$_{p}$ at the convective boundary. D$_{0}$ is the value of the diffusion coefficient determined from the mixing length theory \citep{MLTBohmVitense}.
In this prescription for the CBM, the thermal structure in the mixing zone is that of the envelope, such that $\nabla_{T}$=$\nabla_{rad}$ \citep{Herwig2000}. \Aov{} is linked to convective penetration, also known as ``step-overshoot'', which completely mixes the region with a distance $l$ above the convective boundary given by;
\begin{equation}{
    l = \alpha_{ov} \cdot H_{p}
    }
\end{equation}
Here the thermal structure in the mixing zone is that of the core, such that $\nabla_{T}$=$\nabla_{ad}$ \citep{Zahn1991}. Many different studies attempt to constrain these free parameters, leading to a wide range of possible values, reaching from $\alpha_{ov}$=0.0 \citep{TaoWu2020} to 1.5 \citep{Guenther2014}, depending on the initial mass of the star and whether the effects of rotational mixing were included in the stellar models.\\
\indent Asteroseismology gives us a way to look inside stars and determine the size and mass of their convective cores \citep[see, e.g.,][]{Aerts2003, Straka2005, Briquet2007, Montalban2013, Moravveji2015, Deheuvels2016, Mombarg2019, Angelou2020, Viani2020, Noll2021, Pedersen2021}. These studies, in which the stars do not rotate or rotate at a negligible velocity (at most $\sim$5$\%$ of their critical rotational velocity on the main sequence), revealed that on top of the enhanced CBM (\Aov{}$>$0), mixing in the envelope needs to be included to match asteroseismic observables. Moreover, period-spacing patterns caused by internal gravity waves detected in lightcurves fast rotators reveal that the size of the convective core and the shape of the chemical gradient, on the main sequence is different from what is calculated from most grids of stellar evolution models \citep{Aerts2021}. This leads to the conclusion that the treatment of internal mixing in stellar models needs to be adjusted to bring the models closer to what is observed \citep{Johnston2021}. Similar results are found by other studies, such as \cite{Andrew2020}, based on detached binary systems on the main sequence rather than pulsating stars.\\
\indent \cite{Kaiser2020} did a theoretical study on the effect of the relative importance of convective uncertainties in massive stars, focusing on CBM and the strength of semi-convection. The authors present the stellar structure of models during hydrogen and helium burning, for stars with initial masses of 15, 20, and 25 \msun{}. \cite{Kaiser2020} show that the amount of CBM considered in their models affects the duration of the burning phases, the surface evolution of the stars, the size of the convective cores, but also the isotopic mixture in the core. This has implications for the $^{12}$C/$^{16}$O ratio at the end of core helium burning, which impacts the explosive nucleosynthesis, and thus the supernova yields.\\
\indent In this work, we focus on models with an initial mass of 20 \msun{}, from the main sequence up to a central temperature of 1 GK. 
This mass is chosen because the structure of this star on the main sequence is comparable to one of the most massive gravity-mode pulsators available in asteroseismic studies \citep{Pedersen2021,Pedersen2022}. These pulsators have a convective core and radiative envelope on the main sequence. The determined amount of CBM and envelope mixing, as well as the mixing profile, is similar for stars in the mass range without a strong wind. A 20 \msun{} star is interesting because the star does not undergo strong mass loss during the main-sequence and is expected to explode in a Type-II supernova explosion \citep[see, e.g.,][]{WoosleyHeger2002, Sukhbold2016, LandC2018}. We study the impact of CBM and envelope mixing on the nucleosynthetic wind yields. Also, a 20 \msun{} model is often used in grids of massive stars from various codes, which allows for comparison studies. We vary the CBM and the envelope mixing within the range of values found for $\alpha_{ov}$ and \Denv{} by \cite{Pedersen2021,Pedersen2022} (see Table 2 in the latter work). The models are calculated up to a central temperature of 1 GK to determine the effects on the evolution of the star and on the nucleosynthetic yields.\\
\indent The paper is structured as follows; in Section 2 we describe the physical input for the models and justify the choices made. In Section \ref{Results1} we describe the effects of changing the internal mixing on the stellar evolution, with particular attention to the resulting core composition. In Section \ref{Results3} we discuss the surface abundances and the stellar yields. We finish with a discussion and conclusions in Section \ref{sec:Conclusion}.
\section{Method and input physics}\label{method}
We have used the MESA (Modules for Experiments in Astrophysics) stellar evolution code version 22.11.1 \citep{MESA1,MESA2,MESA3,MESA4, MESA5, MESA6} for the simulations presented in this paper.
\subsection{Nuclear physics input}
\indent We make use of the JINA Reaclib \citep{CyburtJINA2010}. In the latest revisions of MESA, the default reaction rates are set to NACRE \citep{NACRE}, unless not available. The NACRE compilation consists of 86 reaction rates, of which 32 are also contained in the JINA Reaclib. The other 54 rates were replaced by their updated rates as given in version 2.2 of the JINA Reaclib by manually including the tables into MESA. This includes the updated $^{14}$N(p,$\gamma$)$^{15}$O by \cite{Imbriani2005}. This rate has a strong impact on the main-sequence evolution of stars undergoing hydrogen burning via the CNO-cycle.\\
\indent The initial mass of the all models presented here is 20 \msun{}. We use an initial metallicity of Z=0.014 combined with the isotopic mixture as given by \cite{Przybilla2013}. The initial hydrogen content is then defined as X$_{ini}$=1-Y$_{ini}$-Z$_{ini}$. The initial helium content is determined as follows; Y$_{ini}$=0.2465+2.1$\times$Z$_{ini}$, where the value of 0.2465 refers to the primordial helium abundance as determined by \cite{Aver2013}. The value of 2.1 is chosen such that the mass fractions of the chemical mixture adopted from \cite{Przybilla2013} are reproduced \citep[see also][]{Michielsen2021}. Our nuclear network contains all the relevant isotopes for the main burning phases (H, He, C, Ne, O, and Si), allowing us to follow the evolution of the stars in detail up to core collapse. Including the ground and isomeric states of \Al{}, the total nuclear network contains the following 212 isotopes:
n, $ ^{1-3} $H, $ ^{3,4} $He, $ ^{6,7} $Li, $^{7-10}$Be, $ ^{8-11} $B, $ ^{11-14} $C, $ ^{13-16} $N, $ ^{14-19} $O, $ ^{17-20} $F, $ ^{19-23} $Ne, $ ^{21-24} $Na,$ ^{23-27} $ Mg, $ ^{25} $Al, $ ^{26} $Al$ _{g} $,$ ^{26} $Al$ _{m} $,$ ^{27,28} $Al, $ ^{27-33} $Si, $ ^{29-34} $P, $ ^{31-37} $S, $ ^{35-38} $Cl, $^{35-41}$Ar, $^{39-44}$K, $^{39-49}$Ca, $^{43-51}$Sc, $^{43-54}$Ti, $^{47-58}$V, $^{47-58}$Cr, $^{51-59}$Mn, $ ^{51-66} $Fe, $^{55-67}$Co, $^{55-69}$Ni, $^{59-66}$Cu, and $^{59-66}$Zn. This network includes all 204 isotopes that influence the final value of the electron fraction, $Y_{\rm e}$ \citep[see, e.g., ][and references therein]{Farmer2016}, which affects the supernova properties \citep[see][]{Heger2000}.
\subsection{Selection of the mixing parameters}
\indent We make use of the Ledoux criterion to establish the location of the convective boundaries. Convection itself is treated according to the MLT++ prescription \citep{MESA2}. This prescription introduces a free parameter, $\alpha_{mlt}$. Here we set $\alpha_{mlt}$ to 2 for all phases of the evolution.\\
\indent The semi-convection parameter, $\alpha_{sc}$, is set to 0.01 for the entire evolution. Especially for the models with a large amount of CBM, the impact of $\alpha_{sc}$ is limited \citep[see, e.g.,][]{Kaiser2020}. The thermohaline coefficient, $\alpha_{therm}$, is set to 1 for hydrogen and helium burning, and to 0 beyond helium burning to reduce the complexity of the models.
\subsubsection{Convective boundary mixing}
\indent For the main-sequence, we make use of the CBM scheme as implemented by \cite{Michielsen2019}. This scheme allows for a combination of the two CBM schemes mentioned in the introduction, ``exponential'' and ``step-'' overshoot. In the present work, however, we only consider the so-called ``step overshoot'' part of this scheme (also referred to as convective penetration), since most of the B stars analysed by \cite{Pedersen2022} favour this variant for the CBM (55\% versus 45\% for the ``exponential overshoot or diffusive exponentially decaying overshooting''). We vary the CBM, here parametrized by \Aov{}, based on the asteroseismic estimates by \cite{Pedersen2021, Pedersen2022}. For the values for the overshoot parameter, \Aov{}, we take the upper limit for the most massive stars in the sample following the profile where the envelope mixing is governed by internal gravity waves \citep[noted by $\psi_{2,6}$ in][]{Pedersen2022}{}{} either as the most likely candidate or the second most likely candidate, as indicated in Table 6 of \cite{Pedersen2022}. The models with $\psi_{2}$ favour the ``exponential overshoot'' scheme, and for these cases we use a conversion factor of $\sim$10 from the exponential overshoot parameter, $f_{ov}$, to compute the level of the step overshoot parameter such stars would have. This might be a slight underestimation of the conversion, as \cite{ClaretTorres2017} find a factor of \Aov{}/f$_{ov}$=11.36$\pm$0.22, \cite{Moravveji2016} find a factor of 13, and \cite{Valle2017} a factor of 12. Based on a conversion of 10, $\alpha_{ov}$ is varied between 0.05-0.4 on the \textit{main-sequence}, with f$_{0}$=0.005. The parameter f$_{0}$ determines the location inside the convective core from where the CBM starts, which is for computational reasons. This would also be the case for the exponential overshoot scheme \citep[see, e.g.,][Sec. 3.1]{Kaiser2020}.\\
\indent Little is known about the more evolved stages of massive stars from asteroseismology. Therefore, we can look at subdwarf B stars, low mass core-helium burning stars, for guidance. These stars have convective helium-burning cores surrounded by a thin envelope, making their structure similar to that of an evolved massive star, and they are suitable for asteroseismology. Seismic studies of these stars \citep[see, e.g.,][]{VanGrootel2010ApJ, VanGrootel2010AA, Charpinet2010, Charpinet2011, Ghasemi2017, Uzundag2021}, reveal that during the core helium burning stage of these stars additional CBM is necessary to explain the observations. Few pulsating subdwarfs have been modelled, however. Therefore we use a typical amount of CBM with \Aov{}=0.2 for this evolutionary phase and beyond for the convective cores. For hydrogen shell-burning, we again make use of CBM via the ``step-overshoot'' scheme with $\alpha_{ov}$ = 0.20 and f$_{0}$ = 0.005. We did not use CBM for any of the later burning shells aside from the hydrogen-burning shell.
\subsubsection{Enhanced envelope mixing}
\indent Aside from the mixing at the convective boundaries, there is also mixing in the envelopes of a massive star. Extra mixing in the envelope is needed for the stellar evolution models to match the asteroseismic observations. In \cite{Pedersen2021} four different profiles for the envelope mixing are implemented for the asteroseismic modelling; a flat profile leading to a constant envelope mixing, a profile based on hydrodynamic simulations of internal gravity waves \citep{Rogers2017, Varghese2023}, a profile where the mixing is due to vertical shear \citep{Mathis2004}, and lastly a profile where the mixing is due to meridional circulation caused by rotation combined with vertical shear \citep{Georgy2013}. In this work we focus on the profile driven by internal gravity waves because this profile has so far not been used in the global investigation of internal mixing despite that it has been shown to explain asteroseismic data well. Also, this profile has not been investigated beyond the main sequence. The first profile used by \cite{Pedersen2021} cannot satisfactorially reproduce the seismic properties of their sample, and the latter two profiles both are based on the effects of stellar rotation, which have already been extensively investigated in the literature as there are many rotating models available calculated with various codes and implementations \citep[see, e.g.,][]{Heger2000, Ekstrom2008, Ekstrom2012, LandC2013, LandC2018, Banerjee2019, Brinkman2021, MyThesis}. For the implementation of the mixing profile based on internal gravity waves, we follow the implementation of \cite{Michielsen2021}. Here the mixing profile is determined as D$_{IGW}$(r)=D$_{env}$($\rho(r_{CBM})$/$\rho(r)$), where D$_{env}$ is the free parameter and determines the strength of the mixing at the connection point between the extended CBM region and the envelope, $\rho(r_{CBM})$ being the density at the border between the CBM region and the envelope, and $\rho(r)$ is the local density. The value of \Denv{} is varied between 0-6 in steps of 1.5, based on the results by \cite{Pedersen2021, Pedersen2022}. We took the values for \Denv{} from the same models we previously selected for \Aov{}.\\
\indent To determine the location of the convective boundary in the models with limited CBM (\Aov{} = 0.05 and 0.1), we implemented the convective premixing scheme as described in \cite{MESA5} on the main sequence. For this, we enhanced the mesh locally around the convective boundary of the hydrogen burning core.\\
\indent The profile based on internal gravity waves was originally created for the main sequence \citep{Michielsen2021}. Here we also apply this to the rest of the stellar evolution, and most importantly to helium burning. The other burning phases are relatively short and the diffusive mixing is not fast enough to be of significant impact. During helium burning, however, the star has a convective core, which can drive internal gravity waves and thus induce mixing. In the previously mentioned works on subdwarf B stars, no mention is made of enhanced envelope mixing. However, based on the stellar structure, we assume here that the convective helium burning core can drive internal gravity waves as well, and we use the same profile as on the main sequence. The continued enhanced envelope mixing is an important difference from models that consider stellar rotation, since at this point in the evolution, the stellar rotation of the envelope has reduced significantly \citep[see, e.g.,][]{Aerts2019} and its impact on the interior mixing should therefore be diminished.
\subsection{Other input}
\indent The mass-loss scheme is a combination of three different wind prescriptions. For the hot phase (T$_{\rm eff}\geq$ 11kK), we use the prescription given by \cite{Vink2000, Vink2001}, which has been divided by 3 to match the rates of \cite{Bjorklund2021Bistability}. For the cold phase (T$_{\rm eff}\leq$ 1kK) we use \cite{NieuwenhuijzendeJager1990}. For the stars that reach the WR-phase (X$_{surf} \leq$0.4) we use the rates by \cite{NugisLamers2000}. All phases of the wind have a metallicity dependence $\dot{M}\,\propto\,$Z$^{0.85}$ following \cite{Vink2000} and \cite{VinkdeKoter2005}.\\
\indent We have evolved the stars to a central temperature of 10$^{9}$ K, which, depending on the model, is during or after carbon burning. At this point, the mass loss is finished and the wind contribution to the nucleosynthetic yields can be calculated.
\subsection{Yield calculation}
\indent Since our focus is on the pre-supernova nucleosynthetic yields from the winds, we calculate them as follows; we integrate over time the surface mass fraction multiplied by the mass loss. This gives us the total yield of the stellar model. For the stable isotopes, there is a second yield to consider, the net yield, that is, the total yield minus the initial total mass of the isotope originally present in the star. For short-lived radioactive isotopes the net yield is identical to the total yield, because the initial mass present in the stars is zero for these isotopes, and thus no distinction will be made for these yields. Hereafter, the yield will always refer to the total wind yield, unless otherwise indicated.
\section{Stellar models: results and discussion}\label{Results1}
In this section, we discuss the results of our stellar evolution models in terms of evolutionary properties. Table \ref{StellarInfo} contains the key information about the evolutionary stages. 
\subsection{Impact of enhanced mixing on the stellar evolution}
\begin{figure*}
    \centering
    \includegraphics[width=\linewidth]{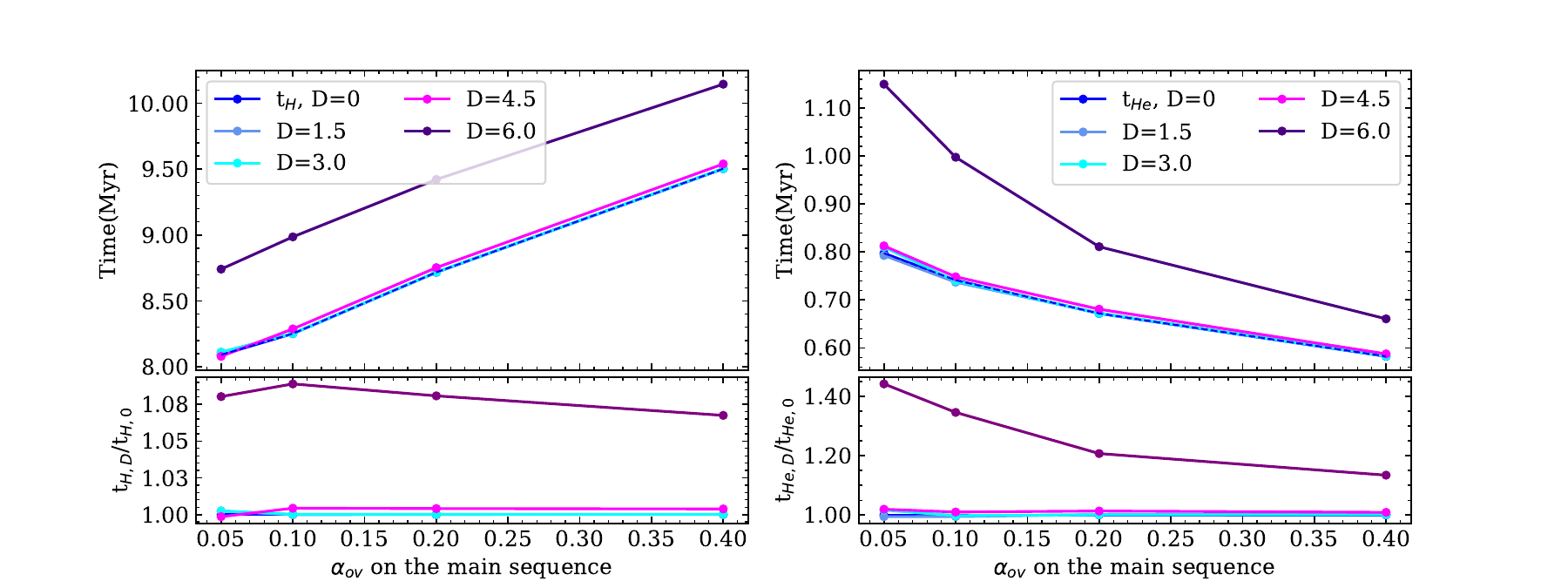}
    \caption{The duration of core hydrogen burning (t$_{\rm H}$, left panel) and core helium burning (t$_{\rm He}$, right panel) as a function of \Aov{}. To highlight the differences, the bottom panels show the ratio between models without additional envelope mixing (reference, in blue) and the other models with the same \Aov{} on the main sequence.}
    \label{Burningtime}
\end{figure*}
\begin{figure*}
    \centering
    \includegraphics[width=0.49\linewidth]{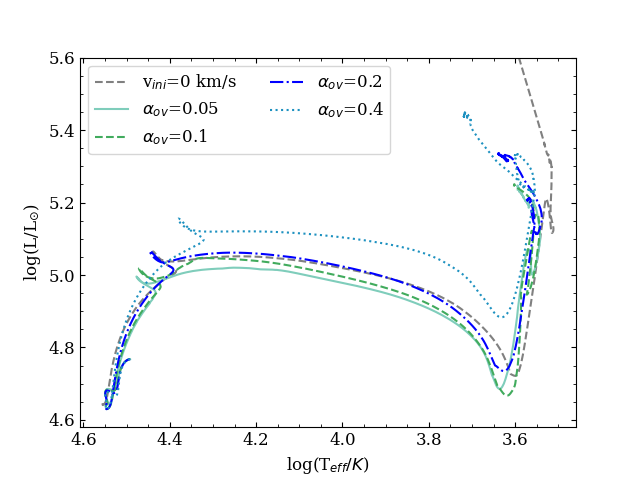}
    \includegraphics[width=0.49\linewidth]{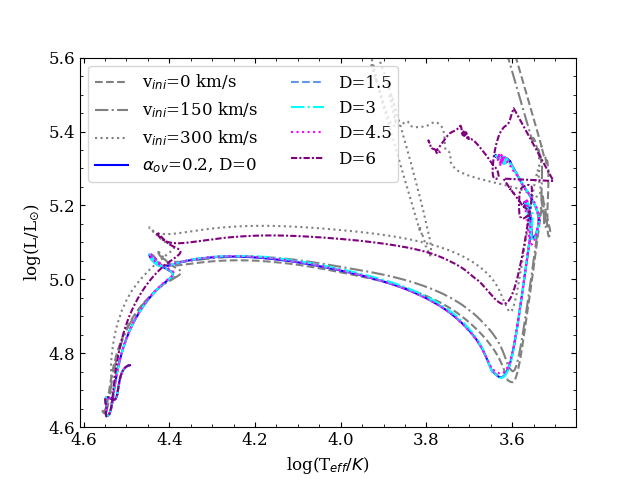}
    \caption{ The tracks in the Hertzsprung-Russel diagram for the different values of \Aov{} and no extra envelope mixing (\Denv{}=0) (left panel) and for the different values of \Denv{} for a fixed value of \Aov{}=0.2 (right panel). The different colours indicate the different values for the CBM and the envelope mixing. In the left panel, the grey line shows the non-rotating model from \cite{Brinkman2021} with \Aov{}=0.2. In the right panel, the rotating models with initial rotational velocities for 150 and 300 km/s are added as a comparison for the different treatment of envelope mixing. The blue line is the same model in both panels. Only the model with \Denv{}=6 shows a significant difference from the other models with the same amount of CBM, while even a moderately different \Aov{} produces a discernible shift.}
    \label{HRDalpha}
\end{figure*}
\begin{figure*}
    \centering
    \includegraphics[width=\linewidth]{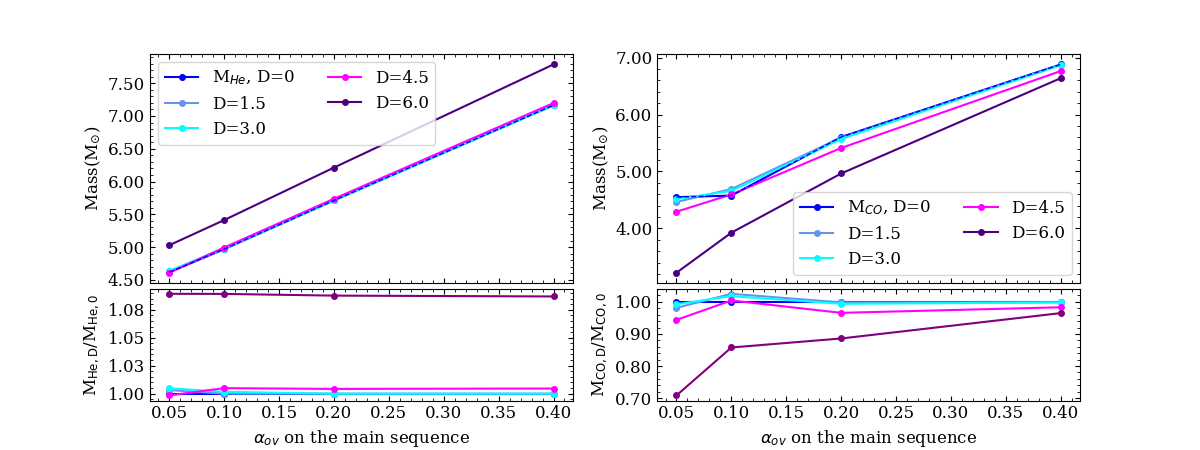}
    \caption{The mass of the stellar cores at the end of hydrogen burning (left panel) and helium burning (right panel) as a function of the overshoot parameter. The lower panels show the ratio between the models with no extra envelope mixing and the other models with the overshoot parameter.}
    \label{CoreMass}
\end{figure*}
\indent On the main-sequence, both the effects of envelope mixing by internal gravity waves and a larger amount of CBM are comparable to the effects of rotational induced mixing. The main-sequence lifetime is extended by increasing values of both \Aov{} and \Denv{} (see left panel of Fig. \ref{Burningtime}). Also, the main-sequence turnoff moves to a higher luminosity and effective temperature, as can be seen in the right panel of Fig. \ref{HRDalpha}. This effect is the strongest for \Denv{}=6, while for the other values of \Denv{} the effect is minimal. These results are in line with the literature for rotational mixing \citep[see, e.g.,][and the grey lines in the right panel of Fig. \ref{HRDalpha}]{Hirschi2004, Ekstrom2012, LandC2018} and for enhanced CBM \citep[see, e.g.,][]{Kaiser2020}. A comparison of rotating versus non-rotating models can be found in, e.g., \cite{Brinkman2021}. In Fig. \ref{HRDalpha} the grey lines show the results of that work, with the model without rotation overlapping mostly with our model with \Aov{}=0.2 and \Denv{}=0. On the evolutionary track, our model with \Denv{}=6 shows a lot of similarities with the model rotating at 300 km/s (see right panel of Fig. \ref{HRDalpha}). Both models have a higher luminosity at the end of the main sequence compared to the other models, and experience more mass loss as well. Both enhanced envelope mixing and enhanced CBM lead to a larger amount of mass loss on the main sequence.\\
\indent After the main sequence, the models with the only enhanced CBM but without enhanced envelope mixing behave comparably to models with rotational mixing, which have shorter helium burning lifetimes compared to their non-rotating counterparts (see right panel of Fig. \ref{Burningtime}). All our models from this point have the same amount of CBM and all differences propagate from differences caused by their the main sequence evolution. This has an analogue with rotational mixing, which loses its efficiency as the rotational velocity of the envelope slows down after the main sequence \citep{Aerts2019}.\\
\indent The models with enhanced envelope mixing are dissimilar from rotating stellar models in the literature and ours without such mixing. The envelope mixing profile based on internal gravity waves requires the existence of a convective core in massive stars, which is still present during the advanced burning stages (except in a few cases where the models undergo radiative carbon burning rather than convective carbon burning, see below). This in principle allows for continued mixing between the envelope and the core as we consider it here. The enhanced mixing affects the size of the helium burning core. At the end of the main sequence the helium cores, hereafter referred as hydrogen-depleted cores to avoid confusion with the helium burning-core, are larger for models with increasing amounts of envelope mixing (see left panel of Fig. \ref{CoreMass}), which is due to an influx of fresh hydrogen from the stellar envelope, not dissimilar to what is happening for models that include rotational mixing.\\
\indent At the onset of helium burning, the models with enhanced envelope mixing, \Denv{}=6.0, have a larger convective core as expected based on the size of their larger hydrogen-depleted cores (see left panel of Fig. \ref{CoreMass}). However, as helium burning proceeds, the convective core region grows less in mass than for the models without enhanced mixing. This is due to the continued influx of helium from the region just above the helium burning core, which then leads to smaller helium depleted cores at the end of helium burning (see column 8 of Table \ref{StellarEvolTable} and the right panel of Fig. \ref{CoreMass}). The models without enhanced mixing need their cores to grow to reach these areas, and a larger hydrogen-depleted core leads to a larger helium-depleted core \citep[see, e.g.,][]{Brinkman2021}. Both the smaller mass of the helium burning core and the larger influx of helium from the region above the core lead to an extended helium burning lifetime (see column 7 of Table \ref{StellarEvolTable}).\\
\indent Overall, the models with the maximum amount of envelope mixing, \Denv{}=6, lose most mass over the whole evolution, increasing with \Aov{} (see column 9 of Table \ref{StellarEvolTable}). For \Aov{}=0.2 and \Aov{}=0.4 and \Denv{}=0-4.5, the total mass loss is comparable, as is the case for \Aov{}=0.05 and \Denv{}=0-4.5. For \Aov{}=0.1, the behaviour of the total mass loss is different as it increases with \Denv{}, though still within 0.9 \msun{} on a total of 14-15 \msun{}. This is due to a slight difference in helium burning, where the convective core remains larger for a higher amount of envelope mixing, leading to a higher luminosity and a higher mass loss rate. This is only the case for the models with \Aov{}=0.1.
\subsection{Impact of enhanced mixing on the core composition}
\begin{figure*}
    \centering
    \includegraphics[width=0.49\linewidth]{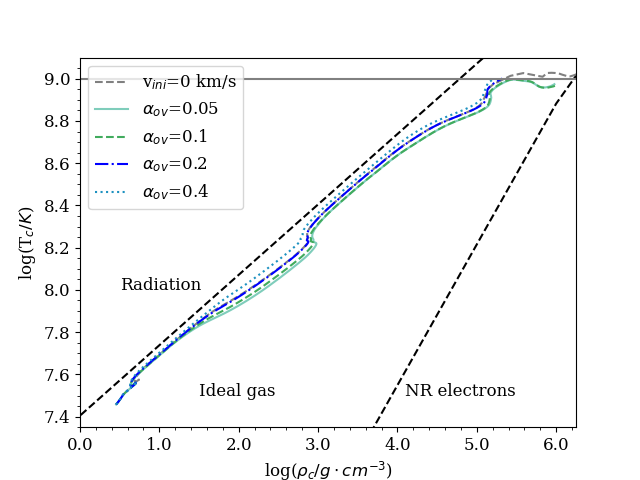}
    \includegraphics[width=0.49\linewidth]{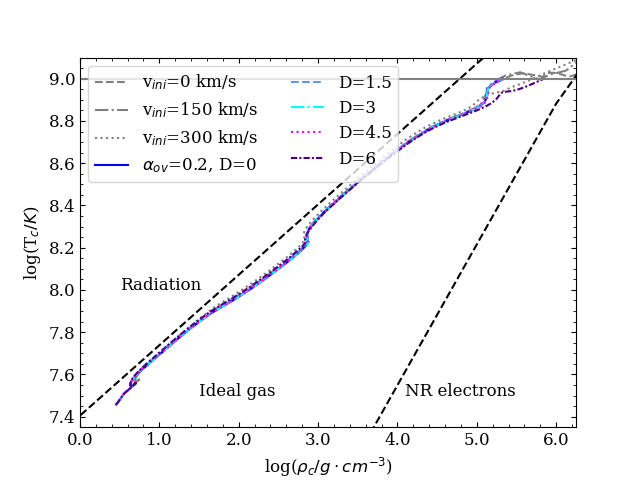}
    \caption{$\rho_{\rm c}$-T$_{\rm c}$ diagram for the 20 \msun{} models with different values for \Aov{} and no extra envelope mixing (\Denv{}=0) (left panel). The right panel shows the same for \Aov{}=0.2 and the different values for the envelope mixing, where D=\Denv{}. In the left panel, the grey line shows the non-rotating model from \cite{Brinkman2021} with \Aov{}=0.2. In the right panel, the rotating models with initial rotational velocities for 150 and 300 km/s are added as a comparison for the different treatment of envelope mixing. The blue line is the same model in both panels. The black dashed lines give a rough indication of the equations of state, i.e., radiative, ideal gas, and non-relativistic electron pressure (NR electrons). The grey horizontal lines indicate a central temperature of T$_{c}$=10$^{9}$.}
    \label{RhocTcAlpha}
\end{figure*}

\indent Fig. \ref{RhocTcAlpha} shows the $\rho_{c}$T$_{c}$-diagrams for the models with different amounts of CBM on the left including the non-rotating model from \cite{Brinkman2021} and different amounts of envelope mixing on the right including the rotating models from \cite{Brinkman2021}. A larger amount of CBM leads to slightly higher core temperatures at a comparable density, i.e., at log($\rho_{c}$/g$\cdot$cm$^{-3}$)=2, log(T$_{c}$/K)=7.93 for \Aov{}=0.05 and log(T$_{c}$/K)=8.00 for \Aov{}=0.4. This is a small difference, but because the strong dependence of the reaction rates on the central temperature, the results can be strong. For example, as a result, the models with \Aov{}=0.05 and 0.1 have already finished carbon burning when they reach log(T$_{c}$/K)=9, while for \Aov{}=0.2 and 0.4 it is still ongoing. For \Denv{}=6.0, the models transition from convective to radiative carbon burning, which leads to a more diagonal track in the diagram. In general, this transition occurs roughly at 22 \msun{} \citep{Heger2000}, which is also found by \cite{Hirschi2004}. In case of the rotating models by the latter authors, it lowers the limit for the formation of a convective carbon-burning core to below 20 \msun{} at an initial rotational velocity of 300 km/s. This is also the case for the model with an initial rotational velocity of 300 km/s from \cite{Brinkman2021}. We find that the enhanced envelope mixing applied is in this sense comparable to the effects of rotational mixing, since both increase the mixing between the top layers of the core and the rest of the envelope.\\
\indent The change from convective carbon burning into radiative carbon burning is due to a change in the mixture of isotopes in the center of the stars, especially in carbon and oxygen. The models that undergo radiative carbon burning have smaller C/O-ratios than their convective counterparts (see column 10 of Table \ref{StellarEvolTable}), by almost a factor 2. This difference is due to the production channels of these isotopes. $^{12}$C and $^{16}$O are both produced during central helium burning, via two sequential reactions. The first reaction is the triple-$\alpha$ reaction which produces $^{12}$C. The second is a subsequent $\alpha$-capture leading to the production of $^{16}$O. Once a high enough abundance of $^{12}$C is reached, the second reaction starts to dominate the energy production. For the models \Denv{}=0, the C/O-ratio decreases with the value for \Aov{}. This is due to the temperature dependence of the reactions, as the more massive cores at the end of the main-sequence have higher central temperatures. The decreasing C/O-ratio with increasing \Denv{} is due to a combination of the extended helium burning lifetime (see column 7 of Table \ref{StellarEvolTable}) and more helium being mixed into the stellar core. A lower carbon content means that less fuel is available for the carbon burning phase ($^{12}$C+$^{12}$C), and thus the star does not form a convective core in this phase.\\
\indent The changes in composition of the stellar core at the end of helium burning due to the enhanced CBM and envelope mixing might have an indirect impact on the explodability of stars. \cite{Chieffi2021} show that a variation the carbon burning regime results in a change of the mass-radius relation \citep[see also][]{Sukhbold2016, LandC2020}, and therefore the compactness parameter, at the time of the core collapse which can potentially impact the final fate of the star \citep{OconnorOtt2011, Ertl2016}. Another potential impact of a change in the core composition is the occurrence of later stage shell mergers, such as between the carbon and oxygen burning shells \citep[see, e.g.,][]{Ritter2018, Andrassy2020, Roberti2024}, though this is beyond the scope of this work.
\section{Wind yields until the end of mass loss}\label{Results3}
In this section we compare the nucleosynthetic wind yields of the various models. We compare the results with those from \cite{Brinkman2021}, which are based on classical rotational mixing instead of asteroseismically calibrated internal mixing as adopted here. We start with three short-lived radioactive nuclei with half-lives less than a Myr; \Al{}, \Cl{}, and \Ca{}. Then we discuss the yields of various stable isotopes; \F{}, \Ne{}, and the isotopes involved in the CNO-cycle, as well as the change in the surface abundance of the stable isotopes.
\subsection{Short-lived radioactive nuclei}
The presence of radioactive isotopes in the early Solar System is well-established as their abundances are inferred from meteoritic data showing excesses in their daughter nuclei \citep[see for a review][]{Lugaro2018}{}{}. Here we discuss three different of such isotopes: \Al{}, a short-lived radioactive isotope with a half life of 0.72 Myr \citep[][]{Alhalflife}, \Cl{} and \Ca{}, with half lives 0.301 Myr \citep[][]{Clhalflife} and 0.0994 Myr \citep[][]{Cahalflife}, respectively. These three radioactive isotopes represent the fingerprint of the local nucleosynthesis that occurred near the time and place of the birth of our Sun. Therefore, these isotopes give us clues about the environment and the circumstances of such birth \citep{adams10, Lugaro2018}. In \cite{Brinkman2021} the effects of stellar rotation on the yields of single stars were investigated. For a 20 \msun{} star with \Aov{}=0.2, the effects of stellar rotation increase the nucleosynthetic yields. Here, we compare the changes in the stellar yields due to CBM and envelope mixing as calibrated by asteroseismology of B stars.
\subsubsection{Aluminium-26}
\begin{figure}
    \centering
    \includegraphics[width=\linewidth]{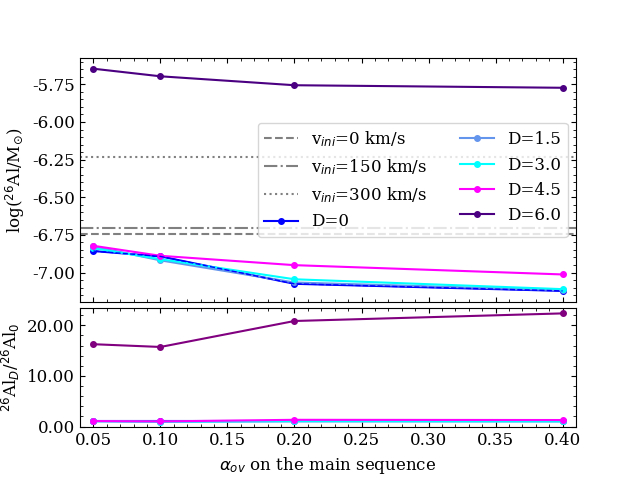}
    \caption{The yields for \Al{} for the different values of \Aov{} and \Denv{} (solid lines) in the upper panel. The horizontal lines indicate the yields of 20 \msun{} models with different rotational velocities (0, 150, and 300 km/s, from \citealt{Brinkman2021}), the dashed, dashed-dotted, and dotted lines, respectively. The lower panel shows the ratio between the yield of the reference models (\Denv{}=0), and the models with the other values for \Denv{}.}
    \label{AlYields}
\end{figure}
\begin{figure*}
    \centering
    \includegraphics[width=0.49\textwidth]{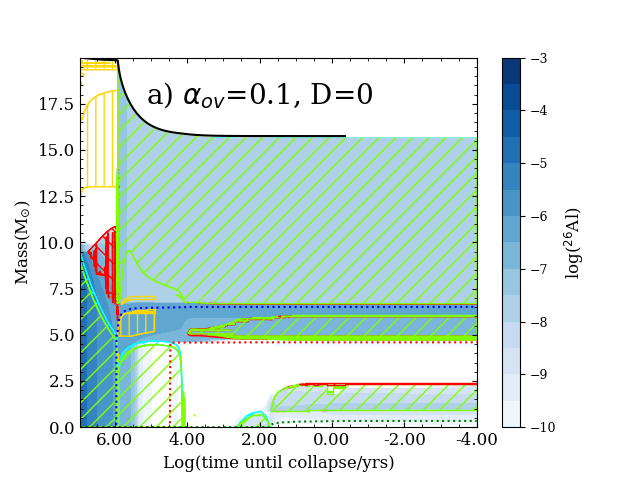}
    \includegraphics[width=0.49\textwidth]{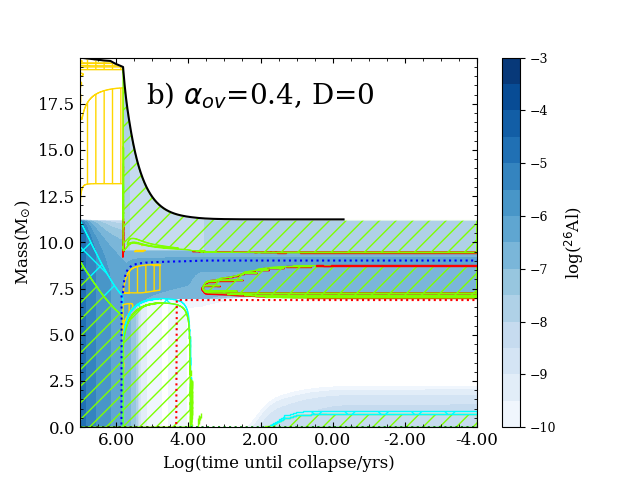}\\
    \includegraphics[width=0.49\textwidth]{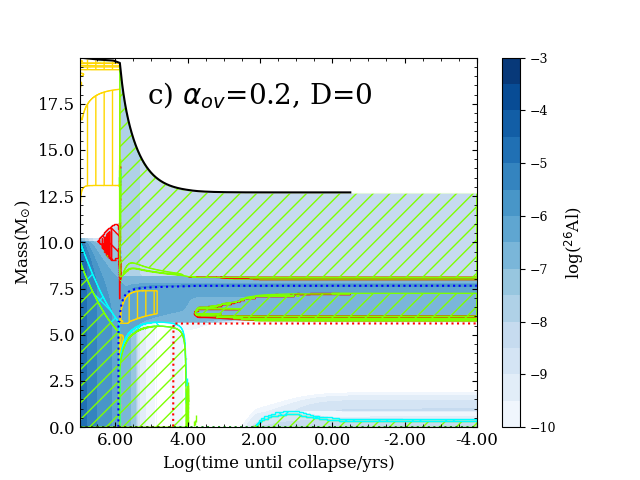}
    \includegraphics[width=0.49\textwidth]{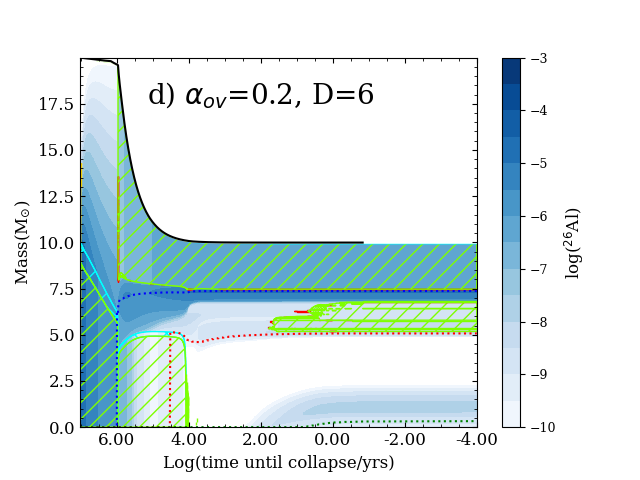}
    \caption{Kippenhahn diagrams for the models with \Aov{} = 0.10, \Aov{}=0.2, and \Aov{}=0,4, all with \Denv{} = 0 (panels a, c, and b respectively), as well as \Aov{}=0.2 and \Denv{} = 6 (panel d). The solid black line is the stellar mass. The green hatched areas correspond to areas of convection, the cyan hatched areas to CBM, the red hatched areas to semi-convection, and the yellow hatched areas to thermohaline mixing. The blue dotted line indicates the hydrogen-depleted core. The red and green dotted lines indicate the helium- and carbon-depleted cores, respectively. The colour scale shows the \Al{} mass fraction as a function of the mass coordinate and time.}
    \label{AlKHDs}
\end{figure*}

\indent \Al{} is a short-lived radioactive isotope which is produced during core and shell-hydrogen burning, during carbon/oxygen convective shell burning, and neon burning during the supernova explosion \citep[see, e.g.,][]{LandC2006}. In this work, we focus on the \Al{} produced in hydrogen burning, which is expelled via the wind. \Al{} is produced by proton captures on the stable isotope $^{25}$Mg. Its main destruction channel during hydrogen burning is the radioactive $\beta$-decay, with a small contribution from the \Al{}(p,$\gamma$)$^{27}$Si. During helium burning, all remaining \Al{} is destroyed via the \Al{}(n,$\alpha$)$^{23}$Na and \Al{}(n,p)$^{26}$Mg. \Al{} is a special isotope since it has an isomeric state, with a half-life of the order of six seconds. The interaction between the ground state and the isomeric state is relatively well understood \citep{iliadis10}, though during hydrogen burning the temperature is such that the two states can be interpreted as separate species \citep[see for recent overviews of \Al{}][]{Diehl2021PASA, Laird2023}.\\
\indent Fig. \ref{AlYields} shows the \Al{} yields of our models. Of the models without envelope mixing, the model with the lowest value for the CBM has the highest yield, even though the mass loss is significantly smaller than for the model with the highest value, i.e., 1.48 \msun{} for \Aov{} = 0.05 versus 9.55 \msun{} for \Aov{} = 0.4. This is due to the difference in the surface mass-fraction of \Al{}, which is lower for the model with $\alpha_{ov}$=0.4. The longer duration of the main sequence for the models with the higher CBM, as described in Section \ref{Results1}, causes the difference in the yields. The difference in the duration of the main sequence, 9.49 Myr for \Aov{}=0.4 versus 8.09 Myr for \Aov{}=0.05, is equal to about twice the half-life of \Al{}. This leads to a larger amount of \Al{} to decay inside the envelope of the star before it can be expelled shortly after the main sequence. Thus, despite the larger mass loss of the models with a higher \Aov{}, this leads to a lower yield for the models with the higher CBM.\\
\indent An increase in envelope mixing has the expected effect on the yields, where more mixing leads to higher yields, especially for the highest value for the envelope mixing, as more processed material is dredged up into the outer envelope. For \Denv{}=6.0 the yield is one order of magnitude larger than for the models without additional mixing. This effect is, to a degree, similar what happens to the yields when rotation is included \citep[see, e.g., Sect. 4.1 of][]{Brinkman2021} which also leads to more mixing between the core and the envelope.\\
\indent In Fig. \ref{AlYields} we also show the yields of the 20 \msun{} models with different rotational velocities (solid, dashed-dotted, and dotted lines). The yields computed in \cite{Brinkman2021} are higher than those of the models presented in this paper, which is due to a difference in the mass-loss scheme. The models computed in \cite{Brinkman2021} lose more mass than those presented here. With less mass lost, our yields are lower for comparable models (\Aov{}=0.2 and \Denv{}=0). Just as increased envelope mixing by internal gravity waves increases the yields, \cite{Brinkman2021} find that increasing the rotational velocity of the stars increases the yields as well, here by a factor of 3.24.\\
\indent Figure \ref{AlKHDs} shows the Kippenhahn diagrams for four of our models with the mass fraction of \Al{} on the colour-scale. At the beginning of the evolution, there is no \Al{} present in the stars. It is produced in the core, and moved to the outer layers by the convective envelope penetrating the area vacated by the retreating hydrogen burning core (a detailed description can be found in \citealt{Brinkman1}), in the case of \Denv{}=0. With higher amounts of \Denv{}, \Al{} is mixed into the envelope at an earlier stage (see panel d of Figure \ref{AlKHDs}.) The model with \Aov{} = 0.1 and \Denv{} = 0 (panel a) has the highest mass fraction of \Al{} on the surface of the star compared to the other models without envelope mixing (panel b and c) by a factor 2.7 and and 3.7, respectively. This leads to the higher yield of this model compared to the other two models with \Denv{} =0 by a factor 1.7 and 1.5, respectively. Panel d shows the model with \Aov{} = 0.2 and \Denv{} = 6.0. The effect of the enhanced envelope mixing is visible in the early disappearance of the thermohaline mixing region (yellow shading) in the envelope, as well as in the earlier upwards diffusion of \Al{}, well before the convective envelope develops, though there is still an abundance gradient. The higher amount of \Al{} in the envelope combined with more mass loss leads to a strong increase in the \Al{} yield (a factor of $\sim$21) compared to the model with \Aov{} = 0.2-\Denv{}=0.
\subsubsection{\Cl{} and \Ca{}}
\begin{figure*}
    \centering
    \includegraphics[width=\linewidth]{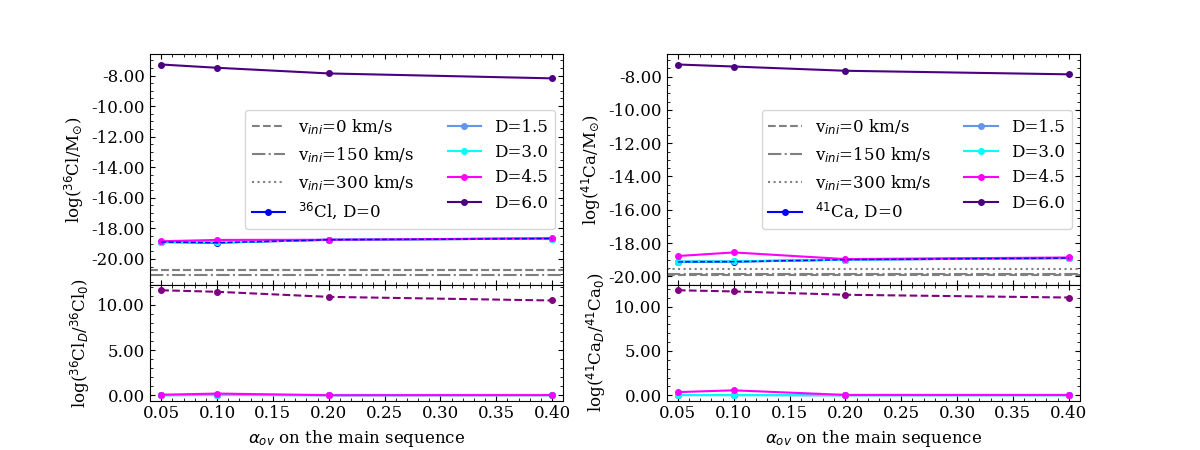}
    \caption{Same as Fig. \ref{AlYields}, but for \Cl{} (left) and \Ca{} (right). The lower panels show the ratio between the yield of the reference models (\Denv{}=0), and the models with the other values for \Denv{}. Mind that unlike in Fig. \ref{AlYields}, here the y-scale for the ratio is in log.}
    \label{ClCaYields}
\end{figure*}

\begin{figure*}
    \centering
    \includegraphics[width=0.49\textwidth]{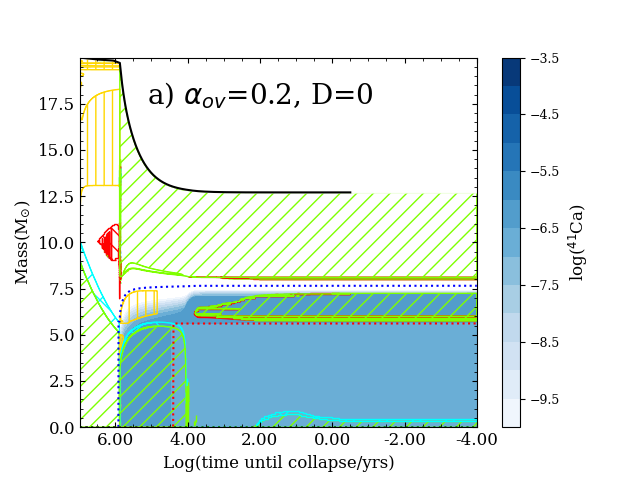}
    \includegraphics[width=0.49\textwidth]{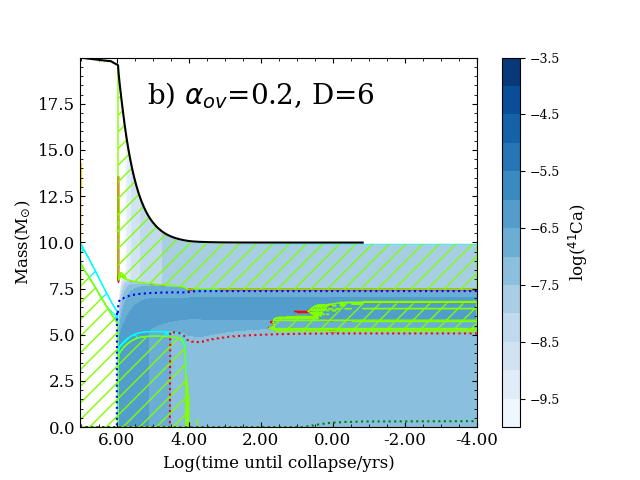}
    \caption{Same as Fig. \ref{AlKHDs}, but with \Ca{} on the colour scale.}
    \label{CaKHDs}
\end{figure*}

\indent \Cl{} and \Ca{} are short-lived radioactive isotopes which are produced during core helium burning and during the explosive burning in a supernova \citep[see, e.g.][]{Pignatari2016, LandC2018}. Both isotopes are mainly produced by neutron captures on the stable isotopes $^{35}$Cl and $^{40}$Ca, respectively. The neutrons needed for this process are produced by the \Ne{}($\alpha,n$)$^{25}$Mg reaction \citep[see, e.g.,][]{Arnould1997}, though in the earliest stages $^{13}$C($\alpha$,n)$^{16}$O has a minor contribution. For both isotopes, the main destruction channel is through further neutron captures, either (n, $\alpha$) or (n,$\gamma$).\\
\indent Fig. \ref{ClCaYields} shows the \Cl{} (left panel) and \Ca{} (right panel) wind yields of the model. For the models with \Denv{}=0-4.5, the yields are very low, all of the order of 10$^{-19}$ (10$^{-20}$) \msun{} for \Cl{} (\Ca{}). This is because \Cl{} and \Ca{} are produced deep inside the stellar interior and need to be brought up to the surface to significantly contribute to the yields. The surface enrichment can be achieved in two manners, (i) by stripping enough mass off the star via the stellar wind to expose the inner parts of the star, and (ii) by diffusing the material produced in the stellar core up to the surface. For the models with \Denv{}=0-4.5, independent of the CBM, the mass loss until the onset of helium burning is not enough to expose the innermost layers of the star, nor is the timescale of helium burning long enough to get any significant diffusion of \Cl{} (\Ca{}) to the stellar surface. For the models with \Denv{}=6.0, the mass loss is also not strong enough to expose the inner layers. However, in this case, the helium burning timescale is long enough for material to be diffused into the envelope, where the convective outer layer brings the material to the surface. This leads to strongly enhanced yields, with a factor 4.18$\cdot$10$^{11}$ (7.35$\cdot$10$^{11}$) to 3.08$\cdot$10$^{10}$ (1.10$\cdot$10$^{11}$) from \Aov{}=0.05 to \Aov{}=0.4. This is visible when comparing the panels of Fig. \ref{CaKHDs}. It shows the Kippenhahn-diagrams for the models with \Aov{}=0.2 and \Denv{}=0 (panel a) and \Denv{}=6.0 (panel b). The colour-scale shows the \Ca{}-mass fraction inside the stars. The distribution of \Cl{} looks similar. For both the model with \Denv{}=0 and \Denv{}=6, the envelope becomes convective after hydrogen burning, and material deeper in the star can be brought to the surface. However, for \Denv{}=0, this does not happen, while for \Denv{}=6.0, the surface becomes enriched in \Ca{} (\Cl{}). This is due to the still-present envelope mixing.\\
\indent When comparing the wind yields for all models with \Denv{}=6.0, they decrease slightly with increasing \Aov{}. This is due to a combination of the duration of the helium burning phase and the mass loss beyond helium burning. The helium burning phase is shorter for the higher values for the CBM, 1.15 Myr for \Aov{}=0.05 versus 0.66 Myr for \Aov{}=0.4, which means that less \Ca{} (or \Cl{}) can reach the surface. Also, the model with \Aov{}=0.05 loses 0.19 \msun{} after helium burning is finished, while the model with \Aov{}=0.4 loses only 0.1 \msun{}. This combination leads to a wind yield of a factor 8.0 (3.9) higher for \Aov{}=0.05 for \Cl{} (\Ca{}).\\
\indent An interesting feature in Fig. \ref{CaKHDs}b is that \Ca{} is visibly mixed out of the core, but only a part reaches the surface of the star. Instead, a band with a higher abundance is formed, roughly at 7.0 \msun{}. Even though the mixing is relatively strong, the helium burning timescale is not long enough to diffuse the produced \Ca{} towards the surface and into the convective layer. This leads to a build-up of \Ca{} above the convective helium-burning core, but below the border of the envelope and the hydrogen-depleted core.\\
\indent Compared to the yields presented here, the \Cl{} and \Ca{} yields from \cite{Brinkman2021} are lower, even for the models that include rotational mixing. This is because the rotational velocity of the envelope has gone down and along with it the efficiency of the mixing in the envelope, while in the asteroseismic models the envelope mixing is still at full strength. Therefore, the surface abundance is lower for the models from \cite{Brinkman2021}, leading to lower wind yields.
\subsection{Stable nuclei}
\subsubsection{\F{} and \Ne{}}
\begin{figure*}
    \centering
    \includegraphics[width=0.49\textwidth]{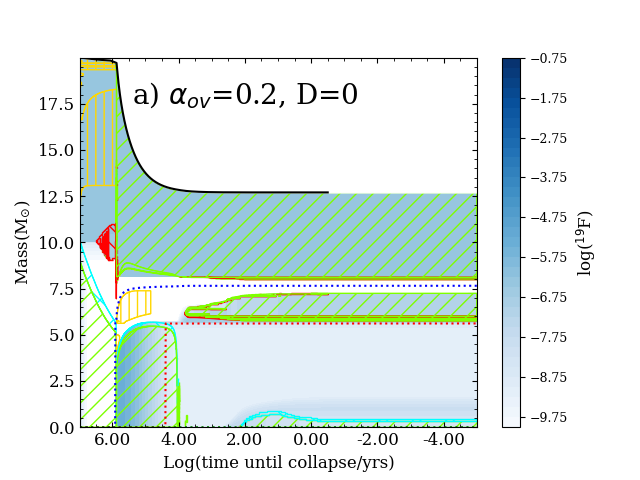}
    \includegraphics[width=0.49\textwidth]{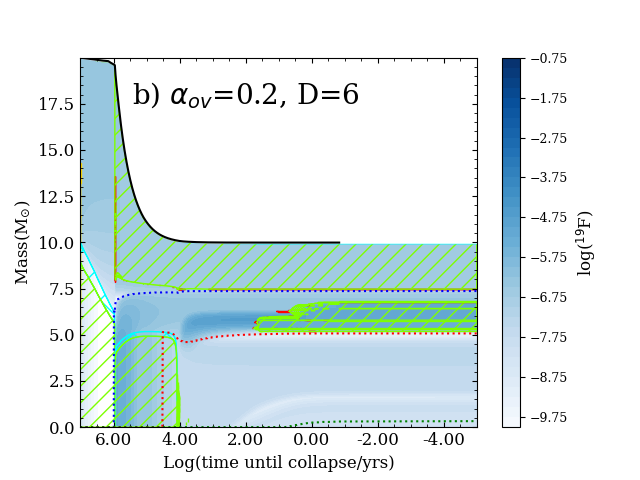}
    \caption{Same as Fig. \ref{AlKHDs} but with \F{} on the colour scale.}
    \label{FKHD}
\end{figure*}

\begin{figure*}
    \centering
    \includegraphics[width=\linewidth]{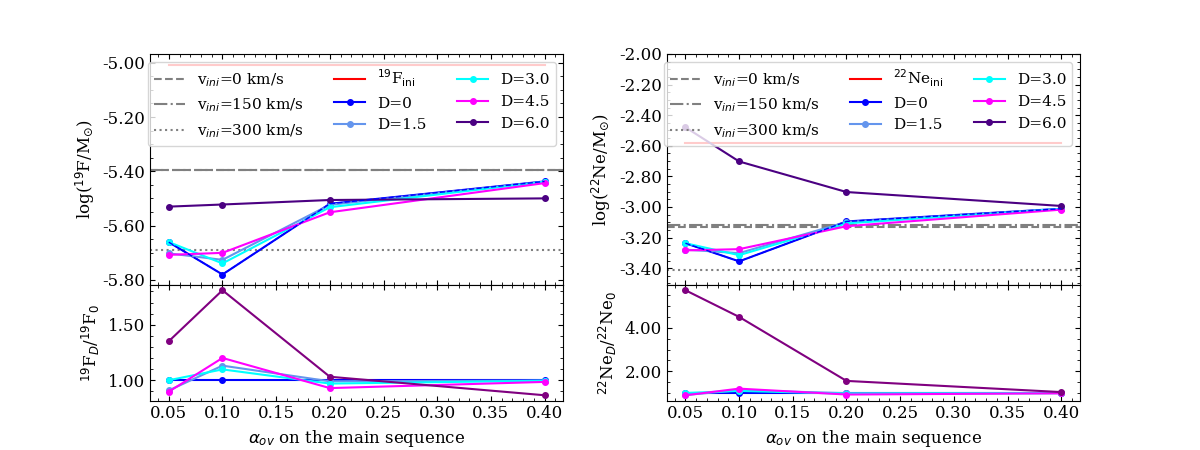}
    \caption{Same as Fig. \ref{AlYields}, but for \F{} (left) and \Ne{} (right). The lower panels show the ratio between the yield of the reference models (\Denv{}=0), and the models with the other values for \Denv{}.}
    \label{FNeYields}
\end{figure*}

\begin{figure*}
    \centering
    \includegraphics[width=0.49\textwidth]{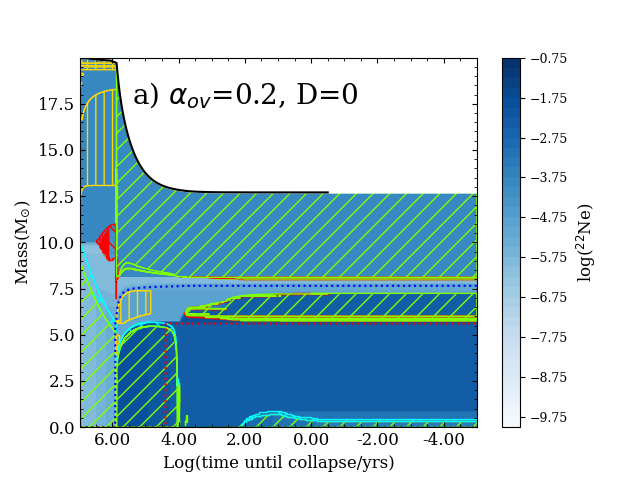}
    \includegraphics[width=0.49\textwidth]{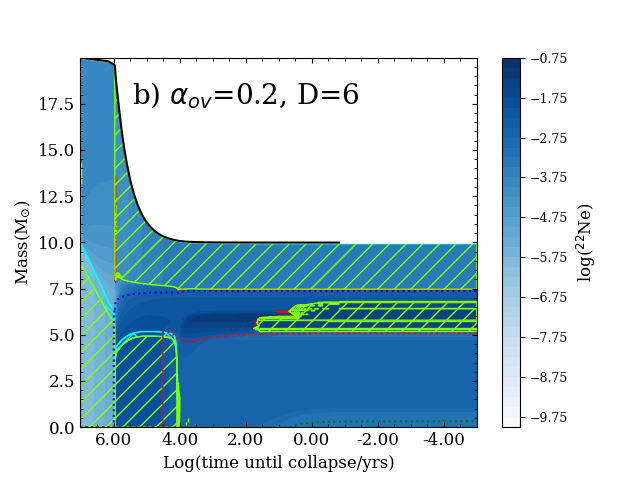}
    \caption{Same as Fig. \ref{AlKHDs} but with \Ne{} on the colour scale.}
    \label{NeKHDs}
\end{figure*}

Both \F{} and \Ne{} are stable isotopes which are present in the stars from birth and their initial abundances therefore scale with the initial metallicity of the star. Fig. \ref{FNeYields} shows the yields for \F{} (left panel) and \Ne{} (right panel). The initial mass of the isotopes present in the stars is given by the red horizontal lines. None of the models presented here have a net positive yield for \F{}, and only one model has a positive yield for \Ne{}.\\
\indent \F\\
\indent \F{} is of interest because its galactic source is still uncertain. \cite{MeynetArnouldF192000} have shown that Wolf-Rayet stars (generally M$_{i} \geq$ 30 \msun{} at solar metallicity) can contribute significantly to the galactic $^{19}$F abundance while \cite{PalaciosF192005} found that the same stars are unlikely to be the source of galactic $^{19}$F, when including updated mass-loss prescriptions and reaction rates. The discussion around $^{19}$F was rekindled by \cite{Joensson2014b,Joensson2014a,Joensson2017} and \cite{Abia2019}, who re-analysed observations of $^{19}$F and proposed that asymptotic giant branch stars are the most likely source of cosmic $^{19}$F. Other works suggest that $^{19}$F may come from rotating massive stars at very low metallicity \citep[see, e.g.,][]{Prantzos2018F, Grisoni2020F, Franco2021F, Roberti2024}. Though, due to the remaining uncertainty in both the mass-loss prescriptions and the reaction rates \citep[see e.g.][]{Stancliffe2005,Ugalde2008}, Wolf-Rayet stars in general cannot be excluded as the sources of galactic $^{19}$F \citep[for a recent overview, see e.g.][]{RydeF192020}. On the other hand, \cite{WomackF19} concluded that Wolf-Rayet stars are not an important contributor. Even though the mass of the models presented here is lower than the Wolf-Rayet limit, the enhanced envelope mixing might have been expected to have an effect.\\
\indent \F{} is typically destroyed in the hydrogen burning core due to proton-captures, via the \F(p, $\alpha$)$^{16}$O reaction, which is part of the CNO-cycles. It is produced during helium burning by a reaction chain starting at $^{14}$N; $^{14}$N($\alpha$,$\gamma$)$^{18}$F(,e$^{+}$)$^{18}$O(p,$\alpha$)$^{15}$N($\alpha$,$\gamma$)$^{19}$F, as first introduced by \cite{Forestini1992F}. It is then rapidly destroyed afterwards due to $\alpha$-captures \citep[][]{MeynetArnouldF192000}{}{}. This is well visible in Fig. \ref{FKHD}, where the colour-scale indicates the mass-fraction of \F{} present in the star. At birth, there is some \F{} present in the star, which is clearly visible in the envelope. In the hydrogen burning core, \F{} is rapidly destroyed. In the left panel (\Denv{}=0), there is no \F{} present below the outermost mass-coordinate of the hydrogen-burning core (10 \msun{}), while for the right panel (\Denv{}=6), the envelope mixing keeps the mass fraction smoothed out in the envelope. During helium burning, which starts at log(time until collapse/yrs)=6.0
in Fig. \ref{FKHD}, \F{} is produced in the stellar core. In the left panel (\Aov{}=0.2 and \Denv{}=0), all the \F{} remains in the core. For the right panel (\Aov{}=0.2 and \Denv{}=6), \F{} is mixed out, as can be seen by the darker shading above the convective helium burning core. Towards the end of helium burning, the central abundance is significantly less, and also in the envelope it is reduced, until the helium burning shell becomes active around log(time until collapse/yrs)=4.0. Panel b of Fig. \ref{FKHD} shows clearly that, despite the enhanced envelope mixing, the \F{} produced in the core is not reaching the surface, which has a comparable mass-fraction over the entire evolution. The \F{} produced in the core is mixed out to roughly the top of the hydrogen-depleted core (blue dotted line in Fig. \ref{FKHD}). The overall effect is that these models have negative \F{} net yields over the whole range (see left panel of Fig. \ref{FNeYields}), as there is more destruction in the star than \F{} leaving the star in the stellar winds.\\
\indent The red line at the top of the left panel of Fig. \ref{FNeYields} is the initial amount of \F{} present in the star. None of the models have positive net yields. Compared to the model with \Denv{}=0, all models with \Denv{}=6 have larger yields, except for \Aov{}=0.4, where the yield is lower. This is because, despite having the largest amount of mass loss, this model also has the longest main sequence lifetime. The strong interaction between the stellar core and stellar envelope leads to an increased destruction of \F{}, which lowers the yield. Also, the shorter duration of helium burning reduces the amount of \F{} mixed into the envelope during helium burning, which together leads to a lower wind yield for this model than for the model without enhanced envelope mixing.The difference in mass lost during helium burning causes the differences in the yields aside from the enhanced destruction. Overall, the yields are fairly comparable, which is mainly because the \F{} expelled was already present in the star from birth and is not newly synthesised.\\
\indent When comparing the yields to those in \cite{Brinkman2021}, we find comparable results for rotational velocities of 0 and 150 km/s, for the lowest envelope mixing. The model with the largest rotational velocity has a lower yield by a factor of $\sim$2, which is due to the destruction of \F{}, enhanced by the induced rotational mixing.\\
\indent \Ne\\
\indent \Ne{} is interesting to study because it is the main neutron sources for the slow neutron-capture process (s-process), via the \Ne{}($\alpha$,n)$^{25}$Mg reaction in massive stars. The amount of \Ne{} and its distribution throughout the star will affect the amount of s-process isotopes produced. Because we consider a star of 20 \msun{}, this will mainly be the weak s-process, which produces the s-process isotopes between iron and strontium, though also slightly beyond that, such as $^{107}$Pd \citep{Pignatari2010}. The extra mixing could potentially produce more \Ne{}, which could lead to a boost in the s-process creating the isotopes beyond the first peak as shown by e.g., \cite{Pignatari2008}, \cite{Frischknecht2012}, \cite{Choplin2018}, \cite{LandC2018}. However, these studies mainly consider low metallicity stars, while we consider solar metallicity.\\
\indent In the right panel of Fig. \ref{FNeYields}, the yields for \Ne{} are shown. For the models with \Denv{}=0-4.5, they follow a similar trend as for \F{}. The \Ne{} wind yields increase strongly for \Aov{}=0.05, with almost a factor 6, while for \Aov{}=0.4, the wind yields barely increase compared to the other models with the same value for the CBM. This is due to a combination of when \Ne{} is produced the difference in the duration of the hydrogen and helium burning phases between the different models, and a difference in the mass loss history.\\
\indent Fig. \ref{NeKHDs} shows the \Ne{}-content for the models with \Aov{}=0.2 and \Denv{}=0 (left) and \Denv{}=6 (right). \Ne{} is partially destroyed during hydrogen-burning in the core by proton-captures via \Ne{}(p,$\gamma$)$^{23}$Na, which is part of the Ne-Na-cycle. Beyond hydrogen burning, \Ne{} is produced again during helium burning, first in the core, and then in the helium burning shell by a chain of $\alpha$-captures on the $^{14}$N left by the CNO-cycle. Part of the \Ne{} produced here is then destroyed by $\alpha$-captures.\\
\indent The strong increase in the wind yield from \Denv{}=0 and \Denv{}=6 for \Aov{}=0.05 is due to the large difference in the mass loss between the two models, 5.2 \msun{} versus 9.1 \msun{}. This is similar for \Aov{}=0.1 and \Aov{}=0.2, though the difference decreases with increasing CBM. For \Aov{}=0.4, the \Ne{} yields are all comparable, which has to do with competing mechanisms. While the mass loss for the model with \Denv{}=6 is the largest with 10.17 \msun{}, the combination of the longest hydrogen burning lifetime with the shortest helium burning lifetime leads to a comparable yield as \Aov{}=0.4 and \Denv{}=0. When the hydrogen burning lifetime increases with more envelope mixing, \Ne{} is mixed from the envelope into the core, where it is destroyed. This reduces the amount of \Ne{} in the envelope and brings the yield down. Also, due to the shorter helium burning lifetime, the \Ne{} produced in the stellar core has limited time to diffuse into the envelope to contribute to the stellar yield. All together, this leads to the models with \Aov{}=0.4 having comparable yields.\\
\indent For the models of \cite{Brinkman2021}, the yields decrease with increasing rotational velocity. This is due to the enhanced mixing between the stellar envelope and the stellar core, just as described above. For the model with an initial rotational velocity of 300 km/s the interaction is the strongest, leading to the lowest yield.
\subsubsection{The stable isotopes of the CNO-cycle}
\begin{figure}
    \centering
    \includegraphics[width=\linewidth]{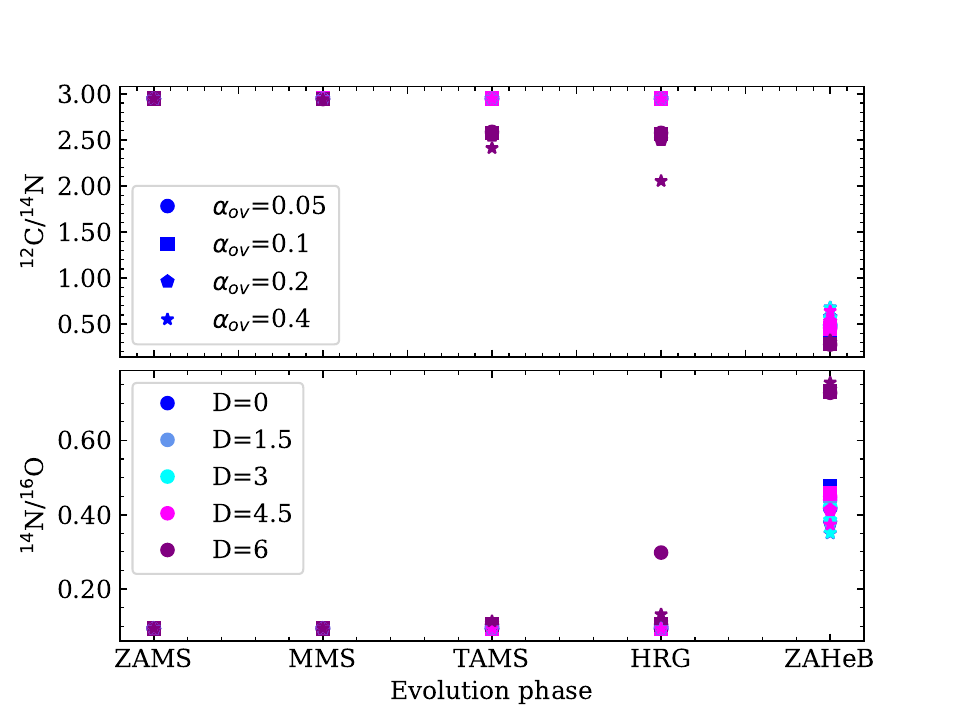}
    \caption{The change of the C/N and N/O-ratios on the surface of the models given at seven different times in the evolution; at the zero-age main-sequence (ZAMS), half-way the main-sequence (MMS), at the terminal-age main-sequence (TAMS), on the Hertzsprung-gap (HRG), and at the onset of helium burning (ZAHeB). The symbols are for the different values for \Aov{} while the colours indicate the different amounts of \Denv{} in the same way as in the previous figures.}
    \label{CNOsurface}
\end{figure}

\begin{figure*}
    \centering
    \includegraphics[width=\linewidth]{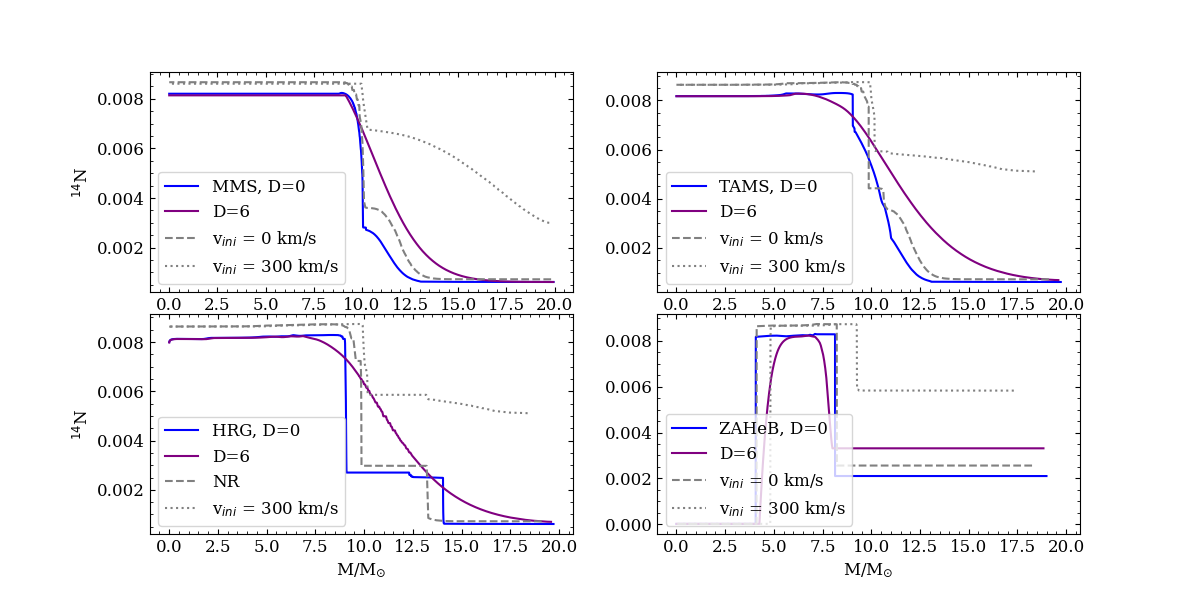}
    \caption{The $^{14}$N abundance profiles at four different phases of the evolution of the models with \Aov{}=0.2 and \Denv{}=0 (blue) and 6 (purple). The top left panel shows the profiles halfway the main-sequence (MMS), the top right at the end of the main sequence (TAMS), the bottom left at the beginning of the Hertzsprung-gap (HRG), and the bottom right at the onset of helium burning (ZAHeB). The grey lines a (dashed for 0 km/s and dotted for 300 km/s) are from \cite{MyThesis}.}
    \label{N14profiles}
\end{figure*}

\indent The isotopes of the CNO-cycle are interesting because they are among the most abundant elements in the Universe. The origin of the CNO-ratios as measured in stardust grains has been extensively studied \citep[see, e.g.,][and references therein]{Romano2003, Romano2019, Kobayashi2020}{}{}. However, their astrophysical sources, especially for carbon, are not well understood \citep[see, e.g.,][]{Kobayashi2020}{}{}. At the same time, due to their large abundance, these elements can be measured in the atmospheres of stars. Especially $^{14}$N has been used as a measure of the rotational mixing \citep[see, e.g.,][]{MandM2000}{}{}. Here we consider the yields of the most abundant isotopes of the CNO-cycle, and briefly discuss how the surface ratios of these isotopes change over the evolution of the star. The solar ratios of the CNO-isotopes are well-known from meteoretic data. While the short-lived radioactive isotopes are produced close in time and location to the early Solar System, the CNO-isotopes have a slow build-up over the galactic history. Therefore, any potential source of the short-lived radioactive isotopes must have a comparable mixture of CNO-isotopes to the early Solar System, or such that the ratios do not change beyond their error-bars \citep[][]{GounelleMeibom2007}{}{}.\\
\begin{figure*}
    \centering
    \includegraphics[width=\textwidth]{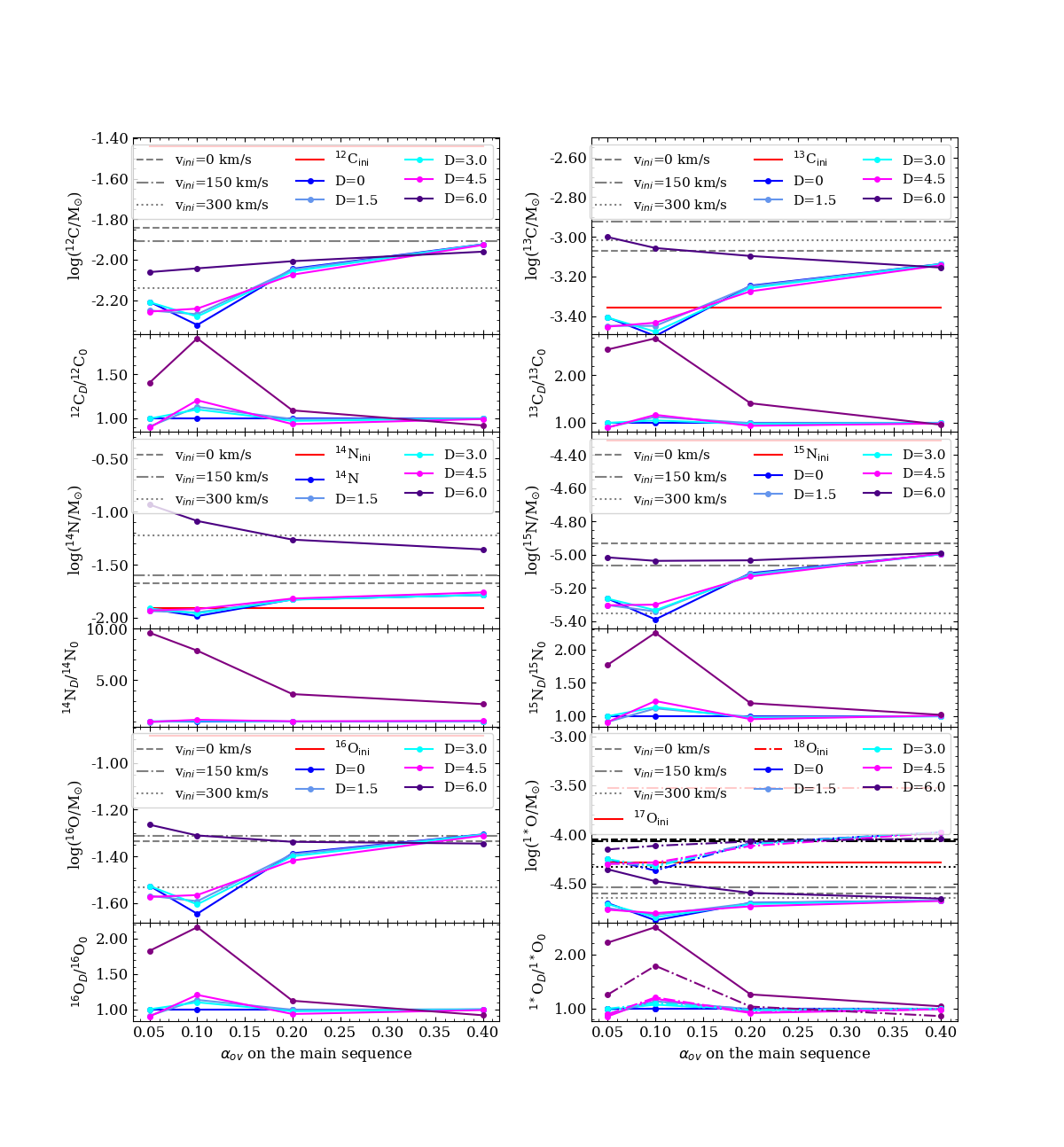}
    \caption{The yields for the main isotopes of the CNO-cycle for the different values of \Aov{} and \Denv{} (dots connected by lines). The horizontal lines indicate the yields of 20 \msun{} models (dashed, dashed-dotted, and dotted lines) with different rotational velocities (0, 150, and 300 km/s, from \cite{MyThesis}). The isotopes are $^{12}$C, $^{13}$C, $^{14}$N, $^{15}$N, $^{16}$O, and in the last panel $^{17}$O (solid lines) and $^{18}$O (dash-dotted lines). In this panel, the grey horizontal lines are for $^{17}$O and the black horizontal lines are for $^{18}$O. The lower panels show the ratio between the yield of the reference models (\Denv{}=0), and the models with the other values for \Denv{}.}
    \label{CNOyields}
\end{figure*}
\indent Fig. \ref{CNOyields} shows the yields of seven isotopes, $^{12,13}$C, $^{14,15}$N, and $^{16-18}$O (the last two isotopes are combined in one panel). In all panels, the initial amount of the isotope present in the models is given by the red line. Overall, the isotopes can be split into three categories, (i) those that are produced and destroyed during helium burning, $^{12}$C and $^{18}$O, (ii) those that are abundantly present at the end of hydrogen burning and are destroyed during helium burning, $^{13}$C and $^{14}$N, and (iii) those that are produced during helium burning and survive the burning stage, $^{15}$N, $^{16}$O, and $^{17}$O. The first group shows a comparable behaviour to \F{}, the second and third group show a comparable behaviour to \Ne{}. All isotopes are initially turned into $^{14}$N as part of the CNO-cycles during core hydrogen burning and then reach an equilibrium value. Nearly all net yields of the isotopes of the CNO-cycle are negative, except for $^{13}$C and $^{14}$N.\\
\indent The CNO-cycle is a catalytic cycle, meaning that as soon as an equilibrium is reached, none of the isotopes involved are overall created or destroyed. The slowest reaction of the CNO-cycle is the $^{14}$N(p,$\gamma$)$^{15}$O, leading to a build-up of $^{14}$N during hydrogen burning. Fig. \ref{CNOsurface} shows how the $^{12}$C/$^{14}$N surface ratio starts to change at the TAMS, mainly for \Denv{}=6.0 (purple symbols). At the onset of helium burning deeper layers of the star, where the composition is changed due to the earlier burning, are exposed due to mass loss, and the ratios have strongly changed compared to the ZAMS. For the $^{14}$N/$^{16}$O-ratio the changes are less visible at the TAMS, and mainly occur at the onset of helium burning. This means that, despite the enhanced mixing in the envelope, the surface of the star is not strongly enriched in processed CNO material during the main sequence.\\
\indent \textit{Nitrogen}\\
\indent Fig. \ref{N14profiles} shows the internal abundance profiles of $^{14}$N at four moments in the stellar evolution. Our models (blue and purple lines) are compared with the non-rotating and fastest rotating models from \cite{Brinkman2021} (grey lines). Halfway up the main-sequence (top left panel) and at the TAMS (top right panel), the model with the highest rotational velocity (dotted line, 300 km/s) has an enhanced surface abundance of $^{14}$N, while the other models do not. It shows that the enhanced envelope mixing smooths out the nitrogen build-up on top of the hydrogen burning core (area up to $\sim$10 \msun{}). The non-rotating model has slightly higher $^{14}$N values just outside of the hydrogen-burning core due the larger value of f$_{0}$, which leads to more mixing from below the convective boundary. On the Hertzsprung-gap (bottom left panel), the semi-convective mixing in the model with \Denv{}=0 causes the step profile while the model with \Denv{}=6.0 has a smooth internal distribution of $^{14}$N. At the onset of helium burning, $^{14}$N is destroyed in the core (below $\sim$ 5 \msun{}). Just outside the helium burning core there is a build-up of $^{14}$N. In the envelope the levels of $^{14}$N are increased due to the hydrogen burning shell, leading to the highest surface mass-fraction for the rotating model.\\
\indent The yields for $^{14}$N and $^{15}$N are given in the central panels of Fig. \ref{CNOyields}. The yields for $^{14}$N follow the same trend as \Ne{}. All yields with \Aov{}=0.2 and 0.4 as well as the models with \Denv{}=6.0 give positive net yields. For the other models the yields are close to the initial amount of $^{14}$N present in the star.\\
\indent \textit{Carbon}\\
\indent The yields for $^{12}$C and $^{13}$C are given in the top panels of Fig. \ref{CNOyields}. During helium burning, $^{12}$C is produced by the triple-$\alpha$ process. Some of this $^{12}$C is then turned into $^{16}$O via subsequent $\alpha$-capture. Due to the combination of production and destruction, the yields of $^{12}$C show a similar behaviour as for \F{}. None of the models presented here have a positive net yield for $^{12}$C. For the models of \cite{MyThesis}, the $^{12}$C yields decrease with increasing initial rotational velocity. As for the previous isotopes, this is due to the enhanced mixing between the core and the envelope. Like for \F{}, the models with \Denv{}=6.0 have increased yields compared to the model without envelope mixing instead of decreasing yields, though at most the yields increase by a factor 1.9.\\
\indent $^{13}$C is completely destroyed by $\alpha$-captures via the $^{13}$C($\alpha$,n)$^{16}$O reaction. This produces the first neutron burst for the production of \Cl{} and \Ca{} at the beginning of helium burning. The $^{13}$C that is expelled via the stellar wind into the interstellar medium, however, is produced during hydrogen-burning as part of the CNO-cycle. It therefore has a similar behaviour as $^{14}$N. $^{13}$C(p,$\gamma$)$^{14}$N is the second-slowest reaction of the CNO-cycle, making $^{13}$C the second most abundant isotope of the cycle. $^{13}$C is produced when the initially more abundant $^{12}$C is converted into $^{13}$N, which then decays into $^{13}$C. Like for $^{14}$N, $^{13}$C is mixed into the envelope, leading to a positive net yield for the models with a strong enough interaction between the core and the envelope (\Aov{}=0.2 and 0.4, and all models with \Denv{}=6.0).\\
\indent The $^{13}$C yields from \cite{MyThesis} increase for an initial rotational velocity of 150 km/s, but then decrease for an initial rotational velocity of 300 km/s. This is because the previously mentioned catalytic nature of the CNO-cycle. The more $^{14}$N is removed from the core, the more $^{13}$C is converted into $^{14}$N, but the compensation is reduced due to the removal of $^{14}$N. This then reduces the yields. In comparison, the models with \Denv{}=6.0 behave more like those with an initial rotational velocity of 150 km/s.\\
\indent \textit{Oxygen}\\
\indent The lower panels of Fig. \ref{CNOyields} give the yields of $^{16-18}$O. $^{16}$O behaves comparable to \Ne{}. However, the initial abundance of $^{16}$O is very high, and due to its late production during helium burning, all the net yields are negative. For $^{17}$O and $^{18}$O, none of the net yields are positive either. $^{17}$O behaves comparable to \Ne{} due to its production in the late stages of core helium burning. All the net yields are negative for this isotope. $^{18}$O is produced as part of the following reaction chain from $^{14}$N into $^{22}$Ne; $^{14}$N($\alpha$,$\gamma$)$^{18}$F(,e$^{+}$)$^{18}$O($\alpha$,$\gamma$)\Ne{}. In the final step, by the end of helium burning, it is almost completely destroyed again. This leads to negative net yields over the whole range of models following the trend of \F{}.
\section{Conclusions}\label{sec:Conclusion}
\indent In this first study of the effect of enhanced CBM and enhanced envelope mixing based on a mixing profile from internal gravity waves, with their strengths based on asteroseismic calibrations, we reach the following conclusions:
\begin{itemize}
    \item On the main sequence, the effects on the stellar evolution are comparable to other studies considering CBM \citep[e.g.,][]{Kaiser2020}, or the effects of rotationally induced mixing \citep[see, e.g.,][and many others]{Ekstrom2012,LandC2018,Brinkman2021}.
    \item During helium burning, enhanced envelope mixing leads to longer helium burning timescales and smaller helium-depleted cores at the end of helium burning. This suggests that while CBM is the dominant source of mixing on the main sequence, where the envelope mixing only has a strong impact for \Denv{}=6, the envelope mixing is the more important mechanism during core helium burning.
    \item The combination of enhanced envelope and CBM has a strong impact on the carbon-to-oxygen-ratio at the end of helium burning. This leads an earlier transition from convective to radiative carbon burning than is seen in the literature. This can potentially have strong effects on the final fate of massive stars and on the explosive nucleosynthesis, as changes in this ratio change the mass-radius relation (compactness) and the potential occurrence of shell mergers.
    \item From \Denv{}=0 to \Denv{}=4.5, the nucleosynthetic yields do not significantly increase. This is because the envelope mixing is a diffusive process and the timescale is too short for significant mixing up to the surface to take place. The main increase happens only for \Denv{}=6, for which the diffusive mixing is strong enough for the isotopes to reach the upper layers of the envelope on the evolutionary timescales of the models. Especially for \Cl{} and \Ca{} the increase is significant with $\sim$ 11 orders of magnitude compared to the models without enhanced envelope mixing.
    \item For \Al{} an increase in the convective boundary leads to lower yields because the longer main sequence lifetimes imply an increased decay of the isotope inside the star.
    \item Out of the stables isotopes, only $^{13}$C and $^{14}$N have mainly positive net yields, and even for those not over the full range of models. For most of the stable isotopes the net yields are negative, as is expected for this mass.
\end{itemize}
We also find that the surface abundance-ratios of our models do not significantly change during the main sequence (see Fig. \ref{CNOsurface}), except for the models with the largest amount of envelope mixing (\Denv{}=6). At the start of helium burning (ZAHeB), all stars show an enhancement of $^{14}$N on the surface, which is not due to envelope mixing, but due to the stellar winds exposing the processed layers of the stellar interior. Our models do not favour a strong nitrogen enrichment of the surface on the main-sequence, even for \Denv{}=6 the enrichment is moderate compared to the rotating model from \cite{MyThesis} (dotted grey line in Fig. \ref{N14profiles}), which is in line with \cite{Aerts2014N14}. In future work we will study the interplay between rotational and internal gravity wave mixing to compute surface abundances and nucleosynthetic yields.\\
\indent Further future work will include, but will not be limited to, the following:
\begin{itemize}
    \item Consideration of different shapes of the mixing profile for the envelopes, based on $\Omega$(r) and d$\Omega$(r)dr profiles, see \cite{Mombarg2022}.
    \item Consideration of time-dependent CBM and time-dependent envelope mixing, see \cite{Varghese2023}.
    \item Inclusion of joint rotational and internal gravtiy wave mixing.
    \item Extension of the current set of models to lower and higher masses to allow for comparison with observations of detached binary systems and comparison with samples of observed stars.
    \item Extension of the models to cover the final phases of the stellar evolution up to core collapse, which can then be used for nucleosynthetic post-processing of the pre-supernova phase to determine the weak s-process yields, as well as the possibility to explode the models and study the impact of envelope mixing based on internal gravity waves on the supernova yields.
\end{itemize}

\begin{acknowledgements}
HEB thanks the MESA team for making their code publicly available, and Ebraheem Farag and Jared Goldberg for their advice on stellar modelling, as well as Rob Farmer for debugging the part of the code related to the reaction rates, and Matthias Fabry for general help with debugging. The research leading to these results has received funding from the KU Leuven Research Council (grant C16/18/005: PARADISE), from the Research Foundation Flanders (FWO) under grant agreement G089422N, as well as from the BELgian federal Science Policy Office (BELSPO) through PRODEX grant PLATO. MM acknowledges support from the Research Foundation Flanders (FWO) by means of a PhD scholarship under project No. 11F7120N.
\end{acknowledgements}

\bibliographystyle{aa} 
\bibliography{references} 

\appendix
\section{Stellar evolution information}\label{StellarEvolTable}
\begin{sidewaystable}[h!]
\caption{Selected properties of the evolution models: M$_{\rm ini}$ is the initial mass in M$_{\odot}$; \Aov{} the amount of CBM on the main sequence; D is short for \Denv{}, the amount of envelope mixing; t$_{\rm H}$, t$_{\rm He}$, and t$_{\rm tot}$ the duration of hydrogen burning, helium burning, and the total evolution time in Myr, respectively. M$_{*,\rm H}$, M$_{*,\rm He}$, and M$_{*,\rm C}$ are the masses of the stars at the end of their respective burning phases in \msun{}; M$_{c,\rm He}$ and M$_{c,\rm C}$ the masses of the hydrogen- and the helium-depleted core, respectively, at the end of the corresponding burning phases in M$_{\odot}$; (C/O)$_{c}$ is the carbon-over-oxygen ratio of the core at the end of helium burning; \Ne{}$_{\rm tot}$ gives the total mass of \Ne{} at the end of helium burning in \msun{}}, $\Delta$M the total mass lost in M$_{\odot}$; T$_{c,f}$ is the final core temperature in K, and $\rho_{c,f}$ is the final core density in g/cm$^{3}$.
\begin{center}\begin{tabular}{ccc|ccc|ccccc|ccc|cc}
\hline 
M$ _{\rm ini} $ & $\alpha_{ov}$ &D& t$ _{\rm H} $ & M$ _{c, \rm He} $ &M$_{*,\rm H}$&t$ _{\rm He} $ & M$ _{c,\rm C} $&M$_{*,\rm He}$ &(C/O)$_{c}$ & \Ne{}$_{\rm tot}$ & M$_{*,\rm C}$ &t$ _{\rm tot} $ &$  \Delta$M &log($\frac{T_{c,f}}{K}$) & log($\frac{\rho_{c,f}}{g*cm^{-3}})$\\
(M$ _{\odot}$) & - &-& (Myr)& (M$ _{\odot} $)& (M$ _{\odot} $)& (Myr)& (M$ _{\odot} $)& (M$ _{\odot} $)&-&(M$ _{\odot} $) & (M$_{\odot}$) & (Myr)& (M$ _{\odot} $) &- & -\\
\hline
20 & 0.05 & 0 & 8.09 & 4.62 & 19.9 & 0.797 & 4.55 & 14.8 & 0.441 &0.0256 & 14.6 & 9.01 & 5.39 & 8.97 & 5.98\\
20 & 0.05 & 1.5 & 8.11 & 4.63 & 19.9 & 0.793 & 4.47 & 15.2 & 0.455& 0.0263& 15 & 9.03 & 4.96 & 8.97 & 6\\
20 & 0.05 & 3 & 8.11 & 4.64 & 19.9 & 0.81 & 4.52 & 14.8 & 0.425 &0.0257& 14.6 & 9.04 & 5.4 & 8.98 & 5.96\\
20 & 0.05 & 4.5 & 8.08 & 4.61 & 19.9 & 0.813 & 4.29 & 15.2 & 0.413 &0.0249 & 15.1 & 9.02 & 4.94 & 8.98 & 5.97\\
20 & 0.05 & 6 & 8.74 & 5.03 & 19.8 & 1.15 & 3.22 & 10.9 & 0.247 &0.146 & 10.7 & 10.0 & 9.27 & 8.84 & 5.53\\
 \hline
20 & 0.1 & 0 & 8.25 & 4.97 & 19.9 & 0.741 & 4.58 & 15.9 & 0.479 & 0.0275& 15.8 & 9.11 & 4.25 & 8.97 & 6.02\\
20 & 0.1 & 1.5 & 8.25 & 4.97 & 19.9 & 0.737 & 4.69 & 15.4 & 0.469 & 0.0273& 15.3 & 9.11 & 4.75 & 8.97 & 5.99\\
20 & 0.1 & 3 & 8.25 & 4.98 & 19.9 & 0.739 & 4.66 & 15.5 & 0.467 &0.0179& 15.4 & 9.11 & 4.61 & 8.97 & 5.98\\
20 & 0.1 & 4.5 & 8.29 & 4.99 & 19.9 & 0.748 & 4.60 & 15.1 & 0.431 &0.0262& 14.9 & 9.15 & 5.06 & 8.98 & 5.95\\
20 & 0.1 & 6 & 8.99 & 5.41 & 19.7 & 0.997 & 3.93 & 10.6 & 0.23&0.0771 & 10.4 & 10.1 & 9.62 & 8.99 & 6.08\\
\hline
20 & 0.2 & 0 & 8.72 & 5.71 & 19.7 & 0.672 & 5.6 & 12.9 & 0.43 &0.0256 & 12.7 & 9.48 & 7.29 & 9 & 5.32\\
20 & 0.2 & 1.5 & 8.72 & 5.71 & 19.7 & 0.673 & 5.59 & 12.9 & 0.431 & 0.0258& 12.8 & 9.48 & 7.24 & 9 & 5.32\\
20 & 0.2 & 3 & 8.72 & 5.71 & 19.7 & 0.672 & 5.57 & 13.0 & 0.43 & 0.0257&12.9 & 9.48 & 7.13 & 9 & 5.32\\
20 & 0.2 & 4.5 & 8.75 & 5.74 & 19.7 & 0.681 & 5.41 & 13.2 & 0.409 &0.025& 13.0 & 9.53 & 6.95 & 9 & 5.33\\
20 & 0.2 & 6 & 9.42 & 6.21 & 19.6 & 0.811 & 4.97 & 10.1 & 0.224 & 0.0362& 9.99 & 10.3 & 10 & 9 & 5.85\\
 \hline	
20 & 0.4 & 0 & 9.50 & 7.16 & 19.5 & 0.582 & 6.88 & 11.4 & 0.403 &0.0235& 11.2 & 10.2 & 8.75 & 9 & 5.22\\
20 & 0.4 & 1.5 & 9.50 & 7.16 & 19.5 & 0.583 & 6.87 & 11.4 & 0.404 &0.0237& 11.3 & 10.2 & 8.74 & 9 & 5.23\\
20 & 0.4 & 3 & 9.50 & 7.16 & 19.5 & 0.583 & 6.87 & 11.4 & 0.403 &0.0236& 11.3 & 10.2 & 8.74 & 9 & 5.24\\
20 & 0.4 & 4.5 & 9.54 & 7.2 & 19.5 & 0.587 & 6.76 & 11.3 & 0.384 &0.0226& 11.2 & 10.2 & 8.79 & 9 & 5.67\\
20 & 0.4 & 6 & 10.1 & 7.79 & 19.3 & 0.66 & 6.64 & 9.83 & 0.232 &0.0198 & 9.73 & 10.9 & 10.3 & 9 & 5.63\\
 \hline	
\end{tabular}\end{center}
\raggedright
\label{StellarInfo}
\end{sidewaystable}
\section{Nucleosynthetic Yields}\label{YieldsTable}

\begin{sidewaystable}
\caption{Wind yields in \msun{} for the models for the isotopes $^{12}$C, $^{13}$C, $^{14}$N, $^{15}$N, $^{16}$O, $^{17}$O, $^{18}$O, \F{}, \Ne{}, \Al{}, \Cl{}, and \Ca{}. \Aov{} is the overshoot on the main-sequence, and \Denv{} the strength of the envelope mixing. For the radioactive isotopes, the yields are not corrected for radioactive decay that might take place in the interstellar medium during the evolution of the star. For the stable isotopes we present the total yields, not corrected for the initial abundance.}
\begin{center}
\begin{tabular}{ccc|cc|cc|ccc|cc|ccc}
\hline 
$M _{\rm ini} $ &  $\alpha_{ov}$ & log(D$_{env}$) & $^{12}$C & $^{13}$C & $^{14}$N & $^{15}$N
& $^{16}$O & $^{17}$O & $^{18}$O &  $^{19}$F & $^{22}$Ne&$^{26}$Al & $^{36}$Cl& $^{41}$Ca\\
(M$ _{\odot}$)& - &- &(M$ _{\odot}$) &(M$ _{\odot}$) &(M$ _{\odot}$) &(M$ _{\odot}$) &(M$ _{\odot}$) &(M$ _{\odot}$) &(M$ _{\odot}$) &(M$ _{\odot}$) &(M$ _{\odot}$) &(M$ _{\odot}$)&(M$ _{\odot}$) &(M$ _{\odot}$) \\
\hline
20 & 0.05 & 0.0 & 0.00618 & 0.000392 & 0.0121 & 5.45e-06 & 0.0298 & 1.99e-05 & 5.58e-05 & 2.18e-06 & 0.00058 & 1.39e-07 & 1.27e-19 & 7.32e-20\\
20 & 0.05 & 1.5 & 0.00561 & 0.000356 & 0.0115 & 4.97e-06 & 0.027 & 1.77e-05 & 5.06e-05 & 1.99e-06 & 0.000526 & 1.49e-07 & 1.25e-19 & 7.7e-20\\
20 & 0.05 & 3.0 & 0.00616 & 0.000391 & 0.0122 & 5.44e-06 & 0.0297 & 1.94e-05 & 5.57e-05 & 2.19e-06 & 0.000579 & 1.45e-07 & 1.3e-19 & 7.64e-20\\
20 & 0.05 & 4.5 & 0.00554 & 0.000352 & 0.0117 & 4.97e-06 & 0.0267 & 1.71e-05 & 4.99e-05 & 1.96e-06 & 0.00052 & 1.51e-07 & 1.42e-19 & 1.65e-19\\
20 &0.05 & 6.0 & 0.00868 & 0.000996 & 0.116 & 9.64e-06 & 0.0544 & 4.4e-05 & 7.03e-05 & 2.95e-06 & 0.00332 & 2.26e-06 & 5.31e-08 & 5.38e-08\\
 \hline											
20 & 0.1 & 0.0 & 0.00476 & 0.000316 & 0.0104 & 4.09e-06 & 0.0226 & 1.33e-05 & 4.28e-05 & 1.66e-06 & 0.000441 & 1.28e-07 & 1.15e-19 & 7.56e-20\\
20 &0.1 & 1.5 & 0.00538 & 0.000355 & 0.0112 & 4.56e-06 & 0.0256 & 1.51e-05 & 4.84e-05 & 1.88e-06 & 0.000499 & 1.2e-07 & 1.26e-19 & 7.89e-20\\
20 & 0.1 & 3.0 & 0.00523 & 0.000332 & 0.011 & 4.64e-06 & 0.0248 & 1.43e-05 & 4.72e-05 & 1.82e-06 & 0.000484 & 1.23e-07 & 1.22e-19 & 7.84e-20\\
20 &0.1 & 4.5 & 0.00573 & 0.000369 & 0.0121 & 5.01e-06 & 0.0272 & 1.58e-05 & 5.16e-05 & 1.99e-06 & 0.00053 & 1.29e-07 & 1.74e-19 & 2.72e-19\\
20 &0.1 & 6.0 & 0.00906 & 0.000877 & 0.0816 & 9.19e-06 & 0.0489 & 3.33e-05 & 7.65e-05 & 3.01e-06 & 0.00198 & 2.01e-06 & 3.24e-08 & 4.07e-08\\
 \hline
20 & 0.2 & 0.0 & 0.00901 & 0.000567 & 0.0149 & 7.76e-06 & 0.041 & 2.01e-05 & 8.15e-05 & 3.02e-06 & 0.000804 & 8.42e-08 & 1.78e-19 & 9.99e-20\\
20 & 0.2 & 1.5 & 0.00894 & 0.000563 & 0.0148 & 7.67e-06 & 0.0407 & 1.99e-05 & 8.08e-05 & 3e-06 & 0.000797 & 8.59e-08 & 1.77e-19 & 9.96e-20\\
20 & 0.2 & 3.0 & 0.00876 & 0.000551 & 0.0148 & 7.54e-06 & 0.0399 & 1.95e-05 & 7.92e-05 & 2.94e-06 & 0.000781 & 9.02e-08 & 1.8e-19 & 1.01e-19\\
20 & 0.2 & 4.5 & 0.00843 & 0.000531 & 0.0152 & 7.41e-06 & 0.0382 & 1.85e-05 & 7.61e-05 & 2.81e-06 & 0.000749 & 1.12e-07 & 1.78e-19 & 1.08e-19\\
20 &0.2 & 6.0 & 0.00983 & 0.0008 & 0.0544 & 9.26e-06 & 0.0459 & 2.53e-05 & 8.47e-05 & 3.12e-06 & 0.00125 & 1.75e-06 & 1.38e-08 & 2.26e-08\\
 \hline	
20& 0.4 & 0.0 & 0.0119 & 0.000729 & 0.0164 & 1.01e-05 & 0.0494 & 2.12e-05 & 0.000105 & 3.65e-06 & 0.000974 & 7.55e-08 & 2.14e-19 & 1.25e-19\\
20 & 0.4 & 1.5 & 0.0119 & 0.000728 & 0.0164 & 1.01e-05 & 0.0493 & 2.12e-05 & 0.000105 & 3.65e-06 & 0.000972 & 7.63e-08 & 2.14e-19 & 1.26e-19\\
20 & 0.4 & 3.0 & 0.0119 & 0.000727 & 0.0164 & 1.01e-05 & 0.0493 & 2.11e-05 & 0.000105 & 3.64e-06 & 0.000971 & 7.76e-08 & 2.15e-19 & 1.27e-19\\
20 & 0.4 & 4.5 & 0.0118 & 0.000719 & 0.0173 & 1.02e-05 & 0.0489 & 2.1e-05 & 0.000104 & 3.6e-06 & 0.000961 & 9.71e-08 & 2.23e-19 & 1.35e-19\\
20 & 0.4 & 6.0 & 0.011 & 0.000699 & 0.044 & 1.03e-05 & 0.0452 & 2.21e-05 & 9.08e-05 & 3.16e-06 & 0.00102 & 1.69e-06 & 6.59e-09 & 1.38e-08\\
    \hline
\end{tabular}\end{center}
\label{BinYields}
\end{sidewaystable}

\end{document}